 \documentclass[10pt,preprint]{aastex}  
 \usepackage{natbib}
\def\ang{\AA}
\def\arcsec{\hbox{$^{\prime\prime}$}}

\def\gapprox{\lower.4ex\hbox{$\;\buildrel >\over{\scriptstyle\sim}\;$}}
\def\lapprox{\lower.4ex\hbox{$\;\buildrel <\over{\scriptstyle\sim}\;$}}

\shortauthors{ASCHWANDEN ET AL 2016}
\shorttitle{VCA-NLFFF Code}

\begin{document}

\title{         The Vertical Current Approximation Nonlinear Force-Free
		Field Code - Description, Performance Tests, and
		Measurements of Magnetic Energies Dissipated in Solar Flares}

\author{        Markus J. Aschwanden$^1$	}

\affil{		$^1)$ Lockheed Martin, 
		Solar and Astrophysics Laboratory, 
                Org. A021S, Bldg.~252, 3251 Hanover St.,
                Palo Alto, CA 94304, USA;
                e-mail: aschwanden@lmsal.com }

\begin{abstract}
In this work we provide an updated description of the {\sl Vertical 
Current Approximation Nonlinear Force-Free Field (VCA-NLFFF)} 
code, which is designed to measure the evolution of the potential,
nonpotential, free energies, and the dissipated magnetic energies 
during solar flares. This code provides a complementary and 
alternative method to existing traditional NLFFF codes. 
The chief advantages of the VCA-NLFFF code over traditional NLFFF
codes are the circumvention of the unrealistic assumption of a 
force-free photosphere in the magnetic field extrapolation method, 
the capability to minimize the misalignment angles between observed 
coronal loops (or chromospheric fibril structures) and theoretical 
model field lines, as well as computational speed. 
In performance tests of the VCA-NLFFF code, by comparing with the 
NLFFF code of Wiegelmann (2004), we find agreement in the potential, 
nonpotential, and free energy within a factor of $\lapprox 1.3$, 
but the Wiegelmann code yields in the average a factor of 2 lower 
flare energies. The VCA-NLFFF code is found to detect decreases 
in flare energies in most X, M, and C-class flares.
The successful detection of energy decreases during a variety
of flares with the VCA-NLFFF code indicates that current-driven
twisting and untwisting of the magnetic field is an adequate
model to quantify the storage of magnetic energies in active 
regions and their dissipation during flares. - The VCA-NLFFF 
code is also publicly available in the {\sl Solar SoftWare (SSW)}. 
\end{abstract}


\section{		INTRODUCTION				}

The measurement of the magnetic field in the solar corona is one of 
the major challenges in solar physics, while measurements of the
photospheric field is a long-standing industry. Some researchers
state that the coronal field cannot be measured (directly),
but we take the standpoint here that a successful modeling method
that matches the observed coronal loop geometries actually equates 
to a real measurement of the coronal magnetic field. The knowledge 
of the coronal magnetic field is paramount in many problems in
solar physics, such as coronal seismology, coronal heating, 
magnetic energy storage, solar wind,
magnetic instabilities, magnetic reconnection, and magnetic
energy dissipation in solar flares and coronal mass ejections, 
which further ties into the global energetics of particle
acceleration and propagation in the coronal and heliospheric 
plasma. While a potential field model is a suitable tool to
explain the approximate geometry of coronal loops, a more
important capability is the nonpotential or free energy, which can be 
liberated in the solar corona and is able to trigger magnetic 
instabilities and to drive eruptive phenomena on the Sun and stars.

Traditional methods compute the magnetic field in the solar
corona by potential-field extrapolation of the photospheric 
line-of-sight component $B_{z}(x,y)$ of the magnetic field, or by 
force-free extrapolation of the photospheric 3D vector field
${\bf B}(x,y)$. Although these methods have been widely and
frequently used in the solar physics community during the last 
three decades, inconsistencies with the observed geometry of 
coronal loops have been noticed recently (Sandman et al.~2009; 
DeRosa et al.~2009), since coronal loops are supposed to accurately
trace out the magnetic field in a low plasma-$\beta$ corona
(Gary 2001). 
Misalignment angles between theoretical {\sl nonlinear force-free
field (NLFFF)} solutions and observed loop directions amount to 
$\mu \approx 24^\circ-44^\circ$ for both potential and nonpotential
field models (DeRosa et al.~2009). 
Several studies have been carried out to pin down the uncertainties
of NLFFF codes, investigating insufficient field-of-views, 
the influence of the spatial resolution,
insufficient constraints at the computation box boundaries, and the 
violation of the force-free assumption in the lower chromosphere
(Metcalf et al.~2008; DeRosa et al.~2009, 2015), but a solution
to reconcile theoretical magnetic field models with the observed
geometry of coronal loops has not been achieved yet and requires 
a new approach. 

Thus there are two different types of NLFFF codes. The first type is
the traditional NLFFF code that uses the 3D vector field 
${\bf B}(x,y)=[B_x(x,y), B_y(x,y), B_z(x,y)]$ from a vector 
magnetograph as input for the photospheric boundary and uses an
extrapolation scheme to compute magnetic field lines in coronal
heights that are consistent with the photospheric boundary condition 
and fulfill the divergence-freeness and the force-freeness conditions. 
Examples and comparisons of such recent NLFFF codes 
are given in Metcalf et al.~(2008) and DeRosa et al.~(2009, 2015),
which include the optimization method (Wheatland et al.~2000;
Wiegelmann 2004; Wiegelmann et al.~2006, 2008; Wiegelmann and Inhester 2010), 
the magneto-frictional method (Valori et al.~2007, 2010), 
the Grad-Rubin method (Wheatland 2007; Amari et al. 2006), 
the conservation-element/solution-element space-time scheme
(CESE-MHD-NLFFF: Jiang and Feng 2013), and other methods. 

The second type is an alternative NLFFF code, which uses only the 
line-of-sight magnetogram $B_z(x,y)$ to constrain the potential field, 
while forward-fitting of an analytical approximation of a special
NLFFF solution in terms of vertical currents to (automatically 
traced) coronal (or chromospheric) loop coordinates $[x(s),y(s)]$ 
is carried out in order to determine the nonlinear force-free 
$\alpha$-parameters for a number of unipolar (subphotospheric) 
magnetic sources. The theory of the vertical-current approximation 
is originally derived in Aschwanden (2013a), while the numeric code 
(called the VCA-NLFFF code here) has been continuously developed and improved
in a number of previous studies (Aschwanden and Sandman 2010;
Sandman and Aschwanden 2011; Aschwanden et al.~2012, 2014a, 2014b;
Aschwanden 2013a, 2013b, 2013c, 2015; Aschwanden and Malanushenko 2013).
Because the recent developments reached now an unprecedented level 
of accuracy in the determination of nonpotential magnetic energies,
it is timely to provide a comprehensive description and performance
tests of the latest version of the VCA-NLFFF method.
A related forward-fitting code, using a quasi-Grad-Rubin method
to match a NLFFF solution to observed coronal loops has been pioneered 
independently (Malanushenko et al.~2009, 2011, 2012, 2014).

In this study we provide a short analytical description of the
VCA-NLFFF method. We start with a description of the organization 
of the code (Section 2), and proceed with the three major parts
of the code: (i) the determination of the potential field (Section 3), 
(ii) the automated tracing of coronal and chromospheric curvi-linear 
structures (Section 4), and the forward-fitting of the VCA-NLFFF model 
to the observed loop geometries (Section 5). Besides brief analytical 
descriptions, we provide extensive performance tests using recently 
published {\sl Atmospheric Imager Assembly (AIA)} (Lemen et al.~2012)
and {\sl Helioseismic Magnetic Imager (HMI)} (Scherrer et al.~2012) data 
from the {\sl Solar Dynamics Observatory (SDO)} (Pesnell et al.~2011),
with particular emphasis on the 
determination of the time evolution of the free (magnetic) energy 
in active regions and the dissipation of magnetic energies during 
X-class flares. Discussions and conclusions are provided in Sections 
6 and 7.

\section{	ORGANIZATION OF THE VCA-NLFFF CODE	} 

In contrast to traditional NLFFF codes, such as the optimization 
method (Wheatland et al.~2000; Wiegelmann and Inhester 2010), 
the magneto-frictional method (Valori et al.~2007, 2010), or 
the Grad-Rubin method (Wheatland 2007; Amari et al. 2006), which
all are designed to match {\sl photospheric} data, the VCA-NLFFF 
code developed here is designed to match {\sl coronal} as well as
{\sl chromospheric} data, a capability that does not exist in
traditional NLFFF codes. The VCA-NLFFF code has a modular
architecture, which can be grouped into three major sections:
(1) The decomposition of a magnetic line-of-sight map into
a number of magnetic charges; (2) the automated feature recognition
of curvi-linear features in coronal (EUV) and chromospheric (UV)
images; and (3) the forward-fitting of a nonpotential magnetic field
model to the observed curvi-linear patterns of coronal
loops or chromospheric fibrils, which we will describe in turn. 
A flow chart of the various modules of the VCA-NLFFF code is 
depicted in Fig.~1.

\subsection{	Time Grid 				}

The initial input for a series of runs is specified by a starting time
$t_{start}$ (in the format of universal time, UT), a desired time interval 
for the entire duration ($t_{dur}$) of magnetic field computations, 
a time cadence ($t_{cad}$),
a heliographic position in longitude $l(t=t_{start})$ and latitude
$b(t=t_{start}$) for the center of 
the chosen field-of-view specified at the starting time, and a desired 
field-of-view (FOV) size for a rectangular subimage (in cartesian
coordinates)
in which a solution of the magnetic field is computed. The subimage 
should be chosen inside the solar disk, because foreshortening near the
limb hampers any type of magnetic modeling. The time input information defines
a time series,
\begin{equation}
	t_i = t_{start} + i \times t_{cad} \ , i=1,...,n_t \ ,
\end{equation}
where the number $n_t$ of time frames amounts to,
\begin{equation}
	n_t={t_{dur} \over t_{cad}} \ ,
\end{equation}
for which magnetic solutions are computed. For flare events we typically
use a cadence of $t_{cad}$=6 minutes (down to 1 min), a duration 
$t_{dur} = \Delta t_{flare} + 2 t_{margin}$ that corresponds to the
flare duration $\Delta t_{flare}$ augmented with a margin of 
$t_{margin}= \pm 1.0$ hrs
before and after the flare event. For the study of the time evolution of
an active region, for instance, we used a total duration of 5 days with 
a cadence of $t_{cad}=6$ minutes, which requires 
$n_t = 5 \times 24 \times 10 = 1200$ time steps (Fig.~2).

The architecture of the VCA-NLFFF code entails sequential processing
of the three major tasks to compute nonpotential field solutions for each
single time step, but an arbitrary number of time steps can be simultaneously
processed in parallel, since the code treats all data from 
each time step independently. The efficient mode of parallel processing 
allows us to compute results in real time, in principle for any time
cadence, if a sufficient number of parallel runs is organized.

\subsection{	Heliographic Coordinates			}

The heliographic position of the center of the chosen FOV subimages is 
automatically updated from the change in heliographic longitude according 
to the differential rotation, 
\begin{equation}
	l(t) = l(t_{start}) + {(t-t_{start}) \over t_{syn}} \times 360^\circ \ ,
\end{equation}
with the synodic period being $t_{syn}(b=0) = 27.2753$ days. The absolute
accuracy of the time-dependent heliographic coordinates is not critical
in the accuracy and continuity of the time-dependent free energy $E_{free}(t)$, 
as long as the chosen field-of-view has weak magnetic fields at the boundaries. 
The relative accuracy among different wavelengths is given by the instrument
cadence, which is $\Delta t_{cad}=12$ s for AIA data from SDO.
Based on this input, the code generates first a catalog with $n_t$ time
entries that contain the times $t_i$ and the time-dependent heliographic
positions $[l_i,b_i]=[l(t_i),b(t_i)]$ for each time step. 

All computations are carried out in a cartesian coordinate system
$[x,y,z]$ with the origin of the coordinate system at Sun center and
$z$ being aligned with the observer's line-of-sight. All solar images
are required to have the solar North-South direction co-aligned with
the $y$-axis (which is provided in the level-1.5 data). The heliographic
positions $[l,b]$ are transformed into cartesian coordinates $[x,y]$
by taking the sinusoidal (annual) variation of the latitude $b_0(t)$ of 
the Sun center into account,
\begin{equation}
	b_0(t) = 7.24^{\circ} \sin{\left[ 2 \pi (t-t_0) \right]} \ ,
\end{equation}
where $t_0$ is the time when the tilt angle of the solar axis is zero
($b_0=0$), when the solar equator coincides with the East-West axis,
which occurs on June 6. From this we can calculate the cartesian
coordinates $[x_0, y_0]$ of a target with heliographic coordinates
$[l, b]$ by,
\begin{equation}
	x_0 = \sin{( l {\pi \over 180^\circ})} \cos{(b {\pi \over 180^\circ})} \ ,
\end{equation}
\begin{equation}
	y_0 = \sin{\left[ ( b-b_0 ) {\pi \over 180^\circ} \right]} \ .
\end{equation}
The left and right-hand side of a chosen square field-of-view with length $FOV$
has then the cartesian coordinates $x_1=x_0-FOV/2$ and $x_2=x_0+FOV/2$, 
while the bottom and top boundaries are $y_1=y_0-FOV/2)$ and $y_2=y_0+FOV/2$.
The relationship between the cartesian and heliographic coordinate systems
is illustrated in Fig.~2 for a time sequence of EUV subimages that follow 
the track of NOAA active region 11158. 

\subsection{	Wavelength Selection 			}

The data input requires a minimum of a line-of-sight magnetogram
and at least one EUV (or UV) image that shows curvi-linear features,
such as coronal loops or chromospheric fibrils. However, an arbitrary
large number of EUV (or UV) images can be supplied for each time step.
The present version of the code deals with HMI/SDO magnetograms
and EUV and UV images from AIA/SDO (94, 131, 171, 193, 211, 304, 335, 1600
\ang ), from the {\sl Interface Region Imaging Spectrograph (IRIS)} 
(De Pontieu et al.~2014) slit-jaw images (1400, 2796, 2832 \ang ), 
from the {\sl Interferometric Bidimensional Spectrometer (IBIS)} 
(Cavallini 2006) (8542 \ang ), or from the {\sl Rapid Oscillations 
in the Solar Atmosphere (ROSA)} (Jess et al.~2010) instrument (6563 \ang ). 
In the present code version, up to 8 
different wavelengths from the same instrument are processed in a single 
run. We will subdivide the wavelengths chosen in each instrument further
by coronal (94, 131, 171, 193, 211, 335 \ang ) and chromospheric 
contributions (304, 1400, 1600, 1400, 2796, 2832, 6563, 8542 \ang ). 
A sample of such a multi-wavelength data set used in magnetic modeling
with the VCA-NLFFF code is depicted in Fig.~3, revealing loop structures 
in coronal wavelengths as well as fibrils and moss in chromospheric
wavelengths, which pose an intricate problem for automated tracing
of magnetic field-aligned curvi-linear features. 

The data input in the VCA-NLFFF code is organized in two ways, either
by reading the data FITS files from a specified directory on the local
computer, or by searching level-1.5 data from local disk cache, or 
remotely from the official SDO and
IRIS data archives (at Stanford University and Lockheed Martin). 
AIA/SDO, HMI/SDO, and IRIS level-1.5 data are available in form of
FITS files with all necessary pointing information, while other
data (e.g., from IBIS and ROSA) need to be prepared with a minimal
FITS header that contains the pointing information (specified with
the FITS descriptors NAXISi, CDELT1i, CRVALi, CRPIXi, and CROTAi,
i=1,2; see Thompson 2006). 

\subsection{	Performance Test Data			}

The validity of the results obtained with the VCA-NLFFF code can
most rigorously be tested with independent data from other NLFFF
codes. The most suitable data set containing published results of 
potential, nonpotential, and free energies has been computed with the 
weighted optimization NLFFF method (Wiegelmann 2004; Wiegelmann 
et al.~2006, 2008; Wiegelmann and Inhester 2010) by Xudong Sun 
for active region AR 11158 during 5 days (2011 February 12-17) 
with a cadence of 12 minutes, yielding 600 time steps for NLFFF
comparisons (Aschwanden, Sun, and Liu 2014b). The corresponding
VCA-NLFFF solutions are calculated for a time cadence of 6 minutes,
yielding 1200 time steps (Fig.~2).

We will use a second test data set of 11 GOES X-class flares
with measurements of potential, nonpotential, and free energies,
computed by Xudong Sun using Wiegelmann's NLFFF method, with a 
12-minute cadence during each flare time interval, providing 
119 time steps for NLFFF comparisons (Aschwanden, Xu, and Jing 2014a). 

A third test data set is used here from the 2014 March 29 GOES X-class
X1.0 flare, the first X-class flare that was observed by IRIS, where
we have simultaneous coverage with AIA and IRIS, so that magnetic
energies based on wavelengths from the coronal as well as chromospheric 
regime can be compared, which was the topic of a previous study 
(Aschwanden 2015). In particular we will use this test data set for a
parametric study of the control parameters operating in the VCA-NLFFF code
(Section 5.6), as listed in Table 1.  

A list of the observing times, time ranges, cadences, number of
time steps, GOES class, heliographic position, NOAA active region
number, and observing instruments of the three test data 
sets is compiled in Table 2.

\section{	THE POTENTIAL FIELD 			}

\subsection{	Analytical Description 			}

The simplest representation of a magnetic potential field 
that fulfills Maxwell's divergence-free condition ($\nabla \cdot {\bf B}=0$)
is a unipolar magnetic charge $m$ that is buried below the solar surface
(Aschwanden 2013a), defining a magnetic field ${\bf B}_m({\bf x})$ that 
falls off with the square of the distance $r_m$,
\begin{equation}
        {\bf B}_m({\bf x})
        = B_m \left({d_m \over r_m}\right)^2 {{\bf r}_m \over r_m} \ ,
\end{equation}
where $B_m$ is the magnetic field strength at the solar surface
above a buried magnetic charge, $(x_m, y_m, z_m)$ is its 
subphotospheric position, $d_m$ is the depth of
the magnetic charge,
\begin{equation}
        d_m = 1-\sqrt{x_m^2+y_m^2+z_m^2} \ ,
\end{equation}
and ${\bf r}_m=[x-x_m, y-y_m, z-z_m]$ is the vector between an arbitrary
location ${\bf x}=(x,y,z)$ in the solar corona 
and the location $(x_m, y_m, z_m)$ of the buried charge.
We define a cartesian coordinate system
$(x,y,z)$ with the origin at the Sun center and 
the direction $z$ chosen along the line-of-sight from Earth
to Sun center. The distance $r_m$ from the magnetic charge is
\begin{equation}
        r_m = \sqrt{(x-x_m)^2+(y-y_m)^2+(z-z_m)^2} \ .
\end{equation}
The absolute value of the magnetic field $B_m(r_m)$ is a function of
the radial distance $r_m$ (with $B_m$ and $d_m$ being constants for a
given magnetic charge),
\begin{equation}
        B_m(r_m) = B_m \left({d_m \over r_m}\right)^2 \ .
\end{equation}
We can generalize from a single magnetic charge to an arbitrary number
$n_{mag}$ of magnetic charges and represent the general magnetic field with a
superposition of $n_{mag}$ buried magnetic charges, so that the potential field
can be represented by the superposition of $n_{mag}$ fields ${\bf B}_m$ from
each magnetic charge $m=1,...,n_{mag}$,
\begin{equation}
        {\bf B}({\bf x}) = \sum_{m=1}^{n_{mag}} {\bf B}_m({\bf x})
        = \sum_{m=1}^{n_{mag}}  B_m
        \left({d_m \over r_m}\right)^2 {{\bf r_m} \over r_m} \ .
\end{equation}
An example of a unipolar charge with a radial magnetic field is a single
sunspot (Fig.~4, top panels), while a dipole field requires two magnetic 
charges with opposite magnetic polarity (Fig.~4, second row). 
Typically, the magnetic field 
of an active region can be represented by a superposition of 
$n_{mag} \approx 20-100$ unipolar magnetic sources, depending on the 
topological complexity of the magnetic field.
Examples of magnetic models with
$n_{mag}=10$ and $n_{mag}=100$ components are shown in Fig.~4 also
(Fig.~4, third and bottom row panels).

A numerical algorithm that deconvolves a line-of-sight magnetogram 
$B_z(x,y,z_{phot})$ has to solve the task of inverting the four observables
$(B_z, \rho_m, z_m, w_m)$ into the four model parameters $(B_m, x_m, y_m, z_m)$
for each magnetic charge, where $B_z$ is the observed line-of-sight component 
of the magnetic field at the photosphere, $\rho_m$ is the apparent distance of 
a magnetic source from Sun center, $z_m=\sqrt{1-\rho_m^2}$ is the distance
from the plane-of-sky (through Sun center), $w_m$ is the FWHM of the magnetic
source, $B_m$ is the magnetic field of the magnetic charge $m$ at the solar
surface, and $(x_m, y_m, z_m)$ are the 3D coordinates of a buried magnetic
charge. The geometric parameters are defined in Fig.~5.
An analytical derivation of the inversion of an observed 
line-of-sight magnetogram into the model parameters of the VCA-NLFFF code
is described in Appendix A of Aschwanden et al.~(2012).
Starting from an approximate initial guess of the aspect angle $\alpha$, 
since $tan{(\alpha)} =(\rho_m/z_m) \approx (\rho_p/z_p)$, we obtain
an accurate value by iterating the following sequence of equations
a few times,
\begin{equation}
        \begin{array}{ll}
        \alpha  &\approx \arctan({\rho_p / z_p}) \\
        \beta_p &=\arctan{\left[ \left( \sqrt{9 + 8 \tan^2 \alpha}-3 \right)
                  / 4\ \tan{\alpha} \right]} \\
        B_m     &={ B_z / [\cos^2{\beta_p} \ \cos{(\alpha-\beta_p)}]} \\
                \beta_2 &=\arccos{\left[
                \left( (\cos{\beta_p})^3 / 2 \right)^{1/3} \right]} \\
        d_m     &={w / \left[ \tan{\beta_2}\ \cos{\alpha} \ (1-0.1\alpha)
                \right]} \\
        r_m     &=(1-d_m)       \\
        \rho_m  &=\rho_p - d_m {\sin{(\alpha-\beta_p)} /
                \cos{\beta_p} } \\
        z_m     &=\sqrt{r_m^2-\rho_m^2} \\
        x_m     &=\rho_m \ \cos{\gamma} \\
        y_m     &=\rho_m \ \sin{\gamma} \\
        \alpha  &=\arctan({\rho_m / z_m}) \\
        \end{array}
\end{equation}
This decomposition procedure of a line-of-sight magnetogram $B_z(x,y)$ 
into a finite number $n_{mag}$ of unipolar magnetic charges, each one
parameterized with 4 parameters $(B_m, x_m, y_m, z_m)$, allows us
to compute the 3D potential field vectors ${\bf B}({\bf x})$ at any 
location of a 3D computation box above the photosphere (with
$r=\sqrt{(x^2+y^2+z^2)} > 1$ solar radius), where the
line-of-sight component $B_z(x,y)$ corresponding to the magnetogram
is just one special component at the curved solar surface,
while the transverse components $B_x(x,y)$
and $B_y(x,y)$ are defined by the same potential model with Eq.~(11).
This algorithm is able to deconvolve magnetograms out to longitudes
of $l \lapprox 80^\circ$ with an accuracy of 
$q_e=E_{model}/E_{obs}=1.000 \pm 0.024$ in the conservation of the
potential energy for a dipolar configuration (see Fig.~21 in 
Aschwanden et al.~2014a), but we limit the application to 
$l \lapprox 45^\circ$ for general magnetograms.

The numerical VCA-NLFFF code contains 4 control parameters (Table 1): 
the number of magnetic charges $n_{mag}$, the width $w_{mag}$ of a 
local map where the magnetic source components are deconvolved, the 
depth range $d_{mag}$ in which magnetic charges are buried, and the
degradation scale $n_{rebin}$ of magnetogram smoothing.
While the number $n_{mag}$ of magnetic charges is a free parameter 
that can be selected by the user, the other control parameters 
were optimized for robustness of results using HMI magnetograms 
(with a pixel size of $0.5\arcsec$), and are set to the constants
$w_{mag}=3$ pixels, $d_{mag}=20$ pixels, and $n_{rebin}=3$ pixels.
The robustness of the results as a function of these control
parameters is also shown in the parameter study in Fig.~15 of 
Aschwanden et al.~(2012). 

\subsection{    Performance Test of Potential Energy  		}

In a first performance test we compare the potential energies
$E_{pot}$ that have been computed simultaneously with the
weighted optimization NLFFF code (Wiegelmann 2004;
Wiegelmann et al.~2006, 2008; Wiegelmann and Inhester 2010),
which we briefly call W-NLFFF in the following, and with our 
VCA-NLFFF code. The potential
energy of an active region or flare is defined here as the
volume integral of the magnetic potential energy integrated
over the entire 3D computation box, with the volume defined
by the chosen FOV in $x$ and $y$-direction, i.e., $[x_1,x_2]$
and $[y_1,y_2]$, while the $z$-range covers the height range 
bound by the photosphere, $z_1(x,y)=\sqrt{(1-x^2-y^2)}$, and 
a curved surface at a height of $h_{max} =0.2$ solar radii, 
i.e., $z_2(x,y)=\sqrt{([1+h_{max}]^2-x^2-y^2}$,
\begin{equation}
	E_{pot} = \int_{x_1}^{x_2} \int_{y_1}^{y_2}
		\int_{z_1(x,y)}^{z_2(x,y)} {B_{pot}^2(x,y,z) \over 8\pi}
		\ dx\ dy\ dz\ \ .
\end{equation}
Note that the total magnetic energy density 
$B^2/8\pi=(B_x^2+B_y^2+B_z^2)/8\pi$
includes both the line-of-sight component $B_z$ and the two
transverse components $B_x$ and $B_y$. The transverse components
are much less accurately known than the line-of-sight component,
by a factor of about 20 for HMI (Hoeksema et al.~2014). The
observed transverse components enter the energy estimate with the
W-NLFFF method, while they are self-consistently determined from
the potential field model with the VCA-NLFFF model, and thus have
a similar accuracy as the line-of-sight component in the
VCA-NLFFF model.

The total potential (or nonpotential) energy is found in a range
of $E_{pot} \approx 10^{32} - 10^{33}$ erg for the active region
NOAA 11158, and increases to $E_{pot} \approx 10^{33} - 10^{34}$ erg 
for X-class flares. 
In Fig.~6 we show the potential energies $E_{pot}^{VCA}$
measured with the VCA-NLFFF code versus the 
potential energies $E_{pot}^W$ measured with the 
W-NLFFF code. Both codes integrate nearly over the same 
volume (although the W-NLFFF code does not take the
sphericity of the solar surface into account), 
but the exact boundaries are not critical since
most of the potential energy comes from the
central sunspot in the FOV area of an active region or flare.
The scatter plots in Fig.~6 show a potential energy ratio of 
$q_{E,pot}=1.22 \pm 0.39$ for 600 measurements of AR 11158
during 5 days observed on 2011 Febr 12-17 
(Aschwanden, Sun, and Liu 2014b), and 	
$q_{E,pot}=0.76 \pm 0.18$ for 119 measurements of 11 X-class
flares observed during 2011-2014 (Aschwanden, Sun, and Liu 2014b).
Thus the average accuracy of the two NLFFF methods agrees within
$\approx 25\%$ for the total potential energy. The accuracy is
similar for the nonpotential energies (Fig.~6, middle panels),
and for the free energy (Fig.~6, bottom panels).
This accuracy is similar to the differences of 12\%-24\% in the
potential energy that was found among different NLFFF codes
for NOAA active region 10978 (DeRosa et al.~2015). 

There are three effects that mostly influence the accuracy of
the potential field measurement, namely the spatial resolution
of the magnetogram, the finite number of magnetic source 
components in the magnetic model, and the asymmetry of sunspots. 
The spatial resolution of a HMI magnetogram is given by the pixel size of
$0.5\arcsec$, which is further downgraded to 
$n_{rebin}=3$ pixels in our VCA-NLFFF code, which prevents a
fragmentation into too many small magnetic elements that do
not significantly contribute to the total energy.
We can measure this effect by calculating the
ratio $q_{E,rebin}$ of potential energies from both the magnetogram
with the original full resolution of $0.5\arcsec$ and the rebinned
magnetogram with $1.5\arcsec$ used in the decomposition of magnetic
sources,
\begin{equation}
	q_{E,rebin} = {\int \int B_{z,rebin}^2(x,y) \ dx\ dy \over
	               \int \int B_{z,full}^2(x,y) \ dx\ dy } \ .
\end{equation}
We find a ratio of $q_{E,rebin}=0.95\pm0.02$ for the potential
energies due to rebinning, for the 1195 time steps of AR 11158 
(Fig.~7, top left panel), and $q_{E,rebin}=0.95\pm0.06$ for 
11 X-class flares (Fig.~7, top right panel). Thus the degradation 
of the magnetogram introduces only an underestimate of $\approx 5\%$
in the potential or nonpotential energy. 

Similarly, we can investigate the effect of the finite number 
$n_{mag}$ of magnetic source components, by calculating the potential
energies from the line-of-sight magnetogram, and by comparing with
the potential energies obtained from the model (with $n_{mag} 
\approx 30$ source components), 
\begin{equation}
	q_{E,model} = {\int \int B_{z,model}^2(x,y) \ dx\ dy \over
	               \int \int B_{z,obs}^2(x,y) \ dx\ dy } \ .
\end{equation}
We find a ratio of $q_{E,model}=0.87\pm0.13$ for the potential
energies, using a model with 30 magnetic sources, 
for the 1195 time steps of AR 11158 
(Fig.~7, bottom left panel), and $q_{E,model}=0.68\pm0.12$ for 
11 X-class flares (Fig.~7, bottom right panel). Thus the model
representation with $\approx 30$ magnetic source components leads
to an average underestimate of about 13\% for magnetic energies 
in active regions, and up to 32\% for the largest flares. The 
optimization NLFFF code of Wiegelmann is found to have a similar
degree of uncertainty, based on the ratio of the total potential 
energy between the model and observed data, i.e.,
$q_{E,model} =1.12-1.24$ (see ratio $E/E_0$ in Table 2 of 
DeRosa et al.~2015).

Further testing revealed that the highest accuracy is not
necessarily controlled by the number of magnetic source components,
although this is true as a statistical trend. Ultimate accuracy
can be achieved when the model parameterization matches the
observed magnetic field distribution, which is fulfilled to the
highest degree for sunspots or magnetic sources with spherical 
symmetry, due to the spherical symmetry of the model definition
of vertical currents in the VCA-NLFFF method (Eq.~11). In contrast, 
asymmetric sunspots require a deconvolution into secondary source 
components, which generally fit the tails of an asymmetric magnetic 
field distribution less accurately than the primary central source
at a local peak in the magnetogram. 

\section{	AUTOMATED LOOP TRACING 				}

\subsection{	General Description 				}

The second major task of the VCA-NLFFF code is the automated tracing of
loop coordinates $[x(s), y(s)]$ (as a function of the loop length coordinate
$s$) in a coronal or chromospheric image, observed
in EUV, UV, optical, or H$\alpha$ wavelengths. The underlying principle of
this task corresponds to convert a 2-D brightness image $(x,y)$ into a
set of 1-D curvi-linear structures $[x(s),y(s)]$. Although there exist a number of
software codes that aim to perform the task of automated pattern recognition
(e.g., see image segmentation methods in Gonzales and Woods 2008), it is our
experience that none of the standard methods yields satisfactory results
for solar data, and
thus we developed a customized code that is optimized for automated detection
of curvi-linear features with relatively large curvature radii (compared with
the width of a loop structure) observed in solar high-resolution images. 

In an initial study, five different numerical codes, designed for automated
tracing of coronal loops in {\sl Transition Region And Coronal Explorer (TRACE)}
images, were quantitatively compared (Aschwanden et al.~2008),
including:
i)  the oriented-connectivity method (OCM), 
ii) the dynamic aperture-based loop segmentation method,
iii) the unbiased detection of curvi-linear structures code,
iv) the oriented-direction method, and the
v) ridge detection by automated scaling method.
One scientific result of this study was that the
size distribution of automatically detected loops follows a cumulative
powerlaw distribution $N(>L) \propto L^{-\beta}$ with $\beta \approx 2.0-3.2$,
which indicates a scale-free process that determines the distribution function 
of coronal loop segments. One of the original five codes was developed further,
a prototype based on the method of {\sl Oriented Coronal CUrved Loop 
Tracing (OCCULT-1)}, which approached an accuracy that was matching visual 
perception of ``hand-traced'' loops (Aschwanden 2010).
An improved code (OCCULT-2) includes
a second-order extrapolation technique (Fig.~8) 
for tracing of curvi-linear features
(Aschwanden et al.~2013a), permitting extended applications to AIA images, 
to chromospheric H$\alpha$ images, as well as applications to images in 
biophysics. While AIA/SDO images have a pixel size of $0.6\arcsec$, 
the automated loop tracing was also extended to higher spatial resolution, 
to images with pixel sizes corresponding to $0.16\arcsec$ (IRIS) and 
$0.1\arcsec$ (IBIS, ROSA) (Aschwanden, Reardon, and Jess 2016). 

The analytical description of the OCCULT-2 code is given in Appendix A.1
of Aschwanden, De Pontieu, and Katrukha (2013a). The IDL source code is
available in the SolarSoftWare (SSW), see the IDL procedure 
{\sl LOOPTRACING$\_$AUTO4.PRO}, and a tutorial is available at the website
{\sl http://www.lmsal.com/$\sim$aschwand} 
{\sl/software/tracing/tracing$\_$tutorial1.html}.

We briefly summarize the numerical algorithm and the control parameters that
can affect the results (see also Table 1). The first step is the background
subtraction, which can be quantified by a minimum level in the original
intensity image ($q_{thresh,1}$), as well as by a minimum level in the
bipass-filtered image ($q_{thresh,2}$). A bipass-filtered image is then
created from a lowpass filter $I_{low}(x,y)$ (i.e., smoothing with a boxcar of 
$n_{sm1}$ pixels) and a highpass filter $I_{high}(x,y)$ (i.e., smoothing 
with a boxcar of $n_{sm2}=n_{sm1}+2$), while the bipass-filter image
$\Delta I(x,y)$ is the difference between the two filters,
\begin{equation}
	\Delta I(x,y) = I_{high}(x,y) - I_{low}(x,y) \ .
\end{equation}
We find that a lowpass filter with $n_{sm1}=1$ and a highpass filter with
$n_{sm2}=n_{sm1}+2=3$ pixels yields best results for most AIA images
(Aschwanden et al.~2013a).
The lowpass filter eliminates large-scale variations in the background,
while the highpass filter eliminates random data noise. For noisy images,
a somewhat higher value is recommended, such as $n_{sm1}=3$ and
$n_{sm2}=n_{sm1}+2=5$. The OCCULT-2 algorithm traces individual
curvi-linear structures by finding first the location ($x_1, y_1$) of the
absolute intensity maximum in the image $\Delta I(x,y)$, then measures the 
direction from the first derivative $dy/dx(x=x_1, y=y_1)$ of the ridge 
that passes true the flux maximum
location ($x_1, y_1$), as well as the curvature radius from the second
derivative $d^2y/dx^2(x=x_1, y=y_1)$ (see geometric diagram in Fig.~8),
and traces the direction of the ridge pixel by pixel, until the end
of the segment traced in forward direction reaches a negative flux
(in the bipass-filtered image). The tracing is then also carried out
in backward direction, in order to find the other end of the loop
segment, yielding the coordinates $[x(s), y(s)]$ of the entire traced loop 
as a function of the loop length coordinate $s$.
The pixels that are located within a half width 
($w_{half}=n_{sm2}/2-1)$ of the ridge coordinates $[x(s), y(s)]$ 
are then erased and the residual
image serves as input for tracing of the second structure,
starting with next flux maximum ($x_2, y_2$), and repeating the
same steps to find the coordinates of the second brightest loop.
The algorithm has 9 control parameters (Table 1), which includes
the maximum number of traced structures $n_{struc}$ per wavelength, 
the lowpass filter $n_{sm1}$, the highpass filter $n_{sm2}=n_{sm1}+2$, 
the minimum accepted loop length $l_{min}$, the minimum allowed
curvature radius $r_{min}$, the field line step $\Delta s$ along
the (projected) loop coordinate, the flux threshold $q_{thresh1}$
in units of the median value of positive fluxes in the original
image), the filter flux threshold $q_{thresh}$ (also in units
of the median value in the positive bipass-filtered fluxes),
and a maximum proximity distance $d_{prox}$ from the location
of the next-located magnetic source. An additional parameter in
the original code is $n_{gap}$, which allowed to skip segments
with negative bipass-filtered fluxes, but is set to $n_{gap}=0$
in the current version of the code. 

\subsection{	Numerical Examples 				}

Examples of bipass-filtered images of all AIA and IRIS (slit-jaw)
wavelengths are shown in Fig.~9, which correspond to the 
observed original images shown in Fig.~3, while the corresponding 
automated loop tracings are shown in Fig.~10, sampled for a 
minimum loop length of $l_{min} \ge 5$ pixels. 

A few special features of the loop tracing code applied to solar
data include the elimination of curvi-linear artifacts resulting
from the boundaries of image portions with saturated fluxes or
pixel bleeding (occurring in CCDs, see AIA 171 and 193 in Fig.~9c and 9d), 
as well as the elimination of strictly horizontal and vertical features, 
that result from edges of incomplete image data, vignetting, or slit markers 
(especially in slit-jaw images from IRIS, see Fig.~9). Scanning through
thousands of images we find many other curvi-linear
features that appear not to be aligned with the magnetic field,
which are eliminated by restricting the maximum
allowed misalignment angle $\mu_2$ iteratively to smaller values
in the forward-fitting algorithm of the VCA-NLFFF code. 

\section{	THE NONLINEAR FORCE-FREE FIELD 			}

\subsection{	Analytical Description 				}

A non-potential field can be constructed by introducing a vertical current
above each magnetic charge, which introduces a helical twist about the
vertical axis (e.g., Fig.~4, top right panel). There is an exact analytical
solution for a straight uniformly twisted flux tube (e.g., Priest 1982),
which can be generalized to the 3-D spherical coordinates of a vertical
flux tube that expands in cross-section according to the divergence-free 
and force-free condition and is accurate to second-order of the force-free 
$\alpha$-parameter (Aschwanden 2013a). This vertical-current approximation
can be expressed by a radial potential field component $B_r(r, \theta)$ and an 
azimuthal non-potential field component $B_{\varphi}(r, \theta)$ in
spherical coordinates $(r, \varphi, \vartheta)$,
\begin{equation}
        B_r(r, \theta) = B_0 \left({d^2 \over r^2}\right)
        {1 \over (1 + b^2 r^2 \sin^2{\theta})} \ ,
\end{equation}
\begin{equation}
        B_\varphi(r, \theta) =
        B_0 \left({d^2 \over r^2}\right)
        {b r \sin{\theta} \over (1 + b^2 r^2 \sin^2{\theta})} \ ,
\end{equation}
\begin{equation}
        B_\theta(r, \theta) \approx 0 \ ,
\end{equation}
\begin{equation}
        \alpha(r, \theta) \approx {2 b \cos{\theta} \over
        (1 + b^2 r^2 \sin^2{\theta})}  \ .
\end{equation}
\begin{equation}
	b = {2 \pi n_{twist} \over l} \ ,
\end{equation}
where $\alpha(r, \theta)$ generally known as force-free $\alpha$-parameter,
is related to the parameter $b$ that expresses the number $n_{twist}$ of 
helical turns over the loop length $l$. The non-potential field of each
magnetic charge $m=1,...,n_{mag}$ can be described with this approximation,
and the associated field components $B_r$ and $B_{\varphi}$ have to be
transformed into a common cartesian coordinate system ${\bf B}_m^{np}({\bf x})$,
and can then be added in linear superposition,
\begin{equation}
	{\bf B}^{np}({\bf x}) = \sum_{m=1}^{n_{mag}} {\bf B}_m^{np} ({\bf x}) \ ,
\end{equation}
which still fulfills the divergence-freeness and force-freeness to second-order
accuracy (Aschwanden 2013a). This way we have a space-filling non-potential
field solution that is parameterized by five variables $(B_m$, $x_m$, $y_m$, 
$z_m$, $\alpha_m$) for each magnetic charge $m=1,..,n_{mag}$, whereof the first four
variables are already determined from the potential-field solution.
From this parameterization we obtain directly a positively defined expression
for the free energy $E_{free}$ (Aschwanden 2013b), which is the difference 
between the non-potential field $E_{NP}$ and the potential field energy $E_P$ 
integrated over the 3-D computation box,
\begin{equation} 
	E_{free} = E_{NP} - E_P = 
	\int {1 \over 8\pi} B_{\varphi}({\bf x})^2 \ q_{iso} \ dV \ .
\end{equation}
where $q_{iso}=(\pi/2)^2 \approx 2.5$ is a correction factor that generalizes
the vertical twist orientation to isotropy (Aschwanden et al.~2014a).

The main task of our VCA-NLFFF code is then to optimize the non-potential
field parameters $\alpha_m, m=1,...,n_{mag}$ by forward-fitting to observed
loop coordinates $[x(s), y(s)]$, which is accomplished by minimizing
the misalignment angles $\mu_{i,j}$ between the theoretical magnetic field model
${\bf B}^{theo}={\bf B}^{np}$ and the observed loop directions ${\bf B}^{obs}$,
\begin{equation}
        \mu_3 = \cos^{-1}\left(
                { ({\bf B}^{\rm theo} \cdot {\bf B}^{\rm obs}) \over
                 |{\bf B}^{\rm theo}| \cdot |{\bf B}^{\rm obs}|} \right) \ ,
\end{equation}
where the 3-D misalignment angles $\mu_{i,j}$ are measured for $i=1,...,n_{loop}$
loops at a number of $j=1,...,n_{seg}$ segments along each loop. The 
optimization criterion minimizes the median of all $\mu_{i,j}$ values.

\subsection{		Numerical Code 	 				}

The numerical implementation of the VCA-NLFFF code has been gradually improved 
over time, including significant changes that are different from earlier numeric
code versions (Aschwanden and Sandman 2010; Sandman and Aschwanden 2011; 
Aschwanden et al.~2012, 2014a, 2014b; Aschwanden 2013b,c, 2015; 
Aschwanden and Malanushenko 2013). 

The VCA-NLFFF code starts with the parameters of the magnetic charges that
were obtained from the potential field fit to the line-of-sight magnetogram
and adds an additional force-free $\alpha$-parameter for each magnetic charge,
so that we have a parameterization of $(x_m, y_m, z_m, B_m, \alpha_m)$,
for $m=1,...,n_{mag}$. On the other side, we have the input of loop
coordinates $[x_{ij}, y_{ij}]$ from $i=1,...,n_{loop}$ loops measured
at $j=1,...,n_{seg}$ segments, where $n_{seg}$ is interpolated to a fixed
number of $n_{seg}=9$ segments, regardless how long the loops are, so that
each loop or loop segment has the same weight in the fitting. Of course,
the third coordinates $z_{ij}$, the line-of-sight coordinates of each loop
are not known a priori, but lower and upper boundaries are given by the
photospheric height $h_{min}=0$ and by the upper boundary of the computation
box at a chosen height of $h_{max}=0.2$ solar radii. In the recent versions 
of the code, an approximate geometry of the height dependence is fitted 
to each loop segment. This approximate geometry encompasses a circular
segment that extends over an arbitrary height range $0 < [h_1, h_2] < h_{max}$,
has a variable orientation of the loop plane, a variable range of curvature 
radii, and covers a variable angular range of a full circle. The loop segment 
can appear as a half circle, a concave or convex circular segment,
of even as a straight line in the extreme limit. A set of such circular
geometries is visualized in Fig.~11. Alternative geometries used in the
parameterization of the height dependence are Bezier spline functions
(Gary et al.~2014). Fitting the 2D projections of this
set of variable circular geometries in height to the 2D projections of the 
observed loop coordinates $[x_{ij}, y_{ij}]$ yields then the best-fit
line-of-sight coordinates $z_{ij}$ for each loop segment, as well as
the vector components of the loop directions at each location $(i,j)$,
\begin{equation}
	  {\bf v}_{ij} = [(x_{i,j+1}-x_{i,j-1}), (y_{i,,j+1}-y_{i,j-1}),
			  (x_{i,j+1}-z_{i,j-1}) ]  \ .
\end{equation}
The code calculates then the misalignment angle $\mu_2$ in 2D (in $x-y$-plane)
and the 3D misalignment angle $\mu_3$ from the scalar product (Eq.~24) between
the observed loop direction ${\bf v}_{ij}$ and the magnetic field vector
${\bf B}_{ij}^{np}$. 
or the potential field (Eq.~11), of a trial nonpotential magnetic field 
(Eq.~17, 18, 22) based on the set of variables $\alpha_m, m=1,...,n_{mag}$.
Thus, in this first iteration step of the forward-fitting procedure we
obtain a set of 3D misalignment angles $\mu_{3,ij}$ for each loop segment,
from which we define an optimization parameter by taking the median of all 
misalignment angles,
\begin{equation}
	\mu_3 = {\rm Median}(\mu_{3,ij}) 
	\ , i=1,...,n_{loop}, \ j=1,...,n_{seg} \ .
\end{equation}
In the second half of an iteration procedure we optimize the global
misalignment angle $\mu_3$ with the minimization procedure of Powell's
method in multi-dimension (Press et al.~1986, p.294), which calculates
in each iteration cycle all gradients $(\partial \mu/\partial \alpha_m),
m=1.,,,,n_{mag}$ of each magnetic charge parameter $\alpha_m$, and improves
the next iteration value by
\begin{equation}
    \alpha_m^{new} = \alpha_m^{old} - \Delta \alpha_0
        {(\partial \mu / \partial \alpha_m) \over 
	max[(\partial \mu/\partial \alpha_m)]} \ ,
\end{equation}
        which optimizes the misalignment angles by
\begin{equation}
        \mu^{new} = \mu^{old} + \Delta \alpha_0 (\partial \mu/\partial \alpha_m) \ ,
\end{equation}
where $\Delta \alpha_0=1.0 \ R_{\odot}^{-1}$ is the maximum increment 
of change in $\alpha_m$ during each iteration step. 
After the first iteration cycle is completed, a second (or subsequent)
cycle is performed, in each one first optimizing the altitudes to obtain
an improved third coordinate $z_{ij}$, and then optimizing the $\alpha_m$.
The final result of a NLFFF solution is contained in a set of
coefficients $(x_m, y_m, z_m, B_m, \alpha_m), m=1,...,n_{mag}$,
from which a volume-filling NLFFF solution
${\bf B}_{np}=[B_x(x,y,z), B_y(x,y,z), B_z(x,y,z)]$ can be computed
in the entire computation box. Individual field lines can be
calculated from any starting point $(x,y,z)$ by sequential
extrapolation of the local B-field vectors in both directions,
until the field line hits a boundary of the computation box.

One of the biggest challenges is the elimination of ``false'' loop tracings, 
which occur due to insufficient
spatial resolution, over-crossing structures, moss structures (Berger et al.~1999;
De Pontieu et al.~1999), data noise, and instrumental effects (image edges, 
vignetting, pixel bleeding, saturation, entrance filter mask, etc.). 
While we attempted to identify such irregularities in earlier code versions,
we find it more efficient to eliminate ``false'' structures iteratively based 
on their excessive misalignment angle. In the present code, ``false'' tracings
are automatically eliminated in each iteration step if they exceed an
unacceptable large value of the misalignment angle. In the latest version
of the code we set a final limit of $\mu_0 \le 20^\circ$ for the
2D misalignment angle, $\mu_2 \le \mu_0$, which is gradually
approached after a sufficient number of iteration steps to warrant convergence.
We set a minimum number of $n_{itmin}=40$ iteration steps, during which the
the maximum acceptable misalignment angle limit $\mu_0$ is linearly
reduced from $\mu_0=90^\circ$ to the final limit of $\mu_0=20^\circ$.
The maximum number of iterations is limited to $n_{itmax}=100$. 

\subsection{    Performance Test : Active Region NOAA 11158 	}

The most extensive data set of potential, nonpotential, and
free energies, computed with both a traditional NLFFF code and
our alternative VCA-NLFFF code is available from NOAA active region 11158,
observed during the 5 days of 2011 Febr 12-17 (Aschwanden, Sun, and Liu 2014b).	
We already compared the potential energies obtained with both codes in Fig.~6.
A time-dependent comparison of the energies obtained with both codes is shown
in Fig.~12. Interestingly, the agreement between the VCA-NLFFF code
the W-NLFFF code is quite good (within $\lapprox 20\%$) during all 5 days,
for the potential (Fig.~12b), the nonpotential (Fig.~12c), as well as for the
free energy (Fig.~12d), which represents a large improvement over previous studies
(see Fig.~8 in Aschwanden, Sun and Liu 2014b), where the free energy
obtained with VCA-NLFFF is substantially lower than the value from W-NLFFF
during all days (June 13-17) except for the first day when the energies
have the lowest values. The good agreement indicates a higher degree
of fidelity for both the VCA-NLFFF and the W-NLFFF code, at least in the
temporal average. The short-term fluctuations are much larger when
modeled with the VCA-NLFFF code, and at this point we cannot discern 
whether the VCA-NLFFF code produces a larger amount of random errors,
or whether the W-NLFFF code produces too much temporal smoothing due
to the preprocessing procedure. 

In Fig.~13 we show the expanded time profile plot that corresponds
to Fig.~9 in Aschwanden, Sun and Liu 2014b). The displayed time profile
of the free energy $E_{free}(t)$ (blue curve in Fig.~13) represents 
the 3-point median values (smoothed with a boxcar of 3 pixels),
which eliminates single-bin spikes (with a
cadence of 6 min here). The time profile $E_{free}(t)$ exhibits much
less random fluctuations than the previous results (Fig.~10 in
Aschwanden, Sun and Liu 2014b), which indicates that the remaining
fluctuations are more likely to be real changes in the free energy.
Indeed, most of the GOES flares show a corresponding dip or decrease
in the free energy, although the detailed timing is not always
strictly simultaneous. In contrast, the time profile of the free
energy $E_{free}(t)$ determined with W-NLFFF shows almost no
temporal fluctuations, and only very small changes or decreases after
a flare, if at all. We suspect that the preprocessing procedure
of the W-NLFFF code over-smoothes changes in the free energy,
and thus underestimates the dissipated energy in a flare. It was
already previously noticed that the W-NLFFF code yields about an order
of magnitude lower energy decreases during flares than the VCA-NLFFF
code (see Fig.~11 in Aschwanden, Sun, and Liu 2014b).

During the observing time interval of 2011 Feb 12-17, a total of 36
GOES C, M, and X-class flares were identified in the NOAA flare catalog.
For most of these 36 events we see a significant decrease of free energy
with the VCA-NLFFF method. We show 9 examples with the most significant
energy decreases in Fig.~14, for both the VCA-NLFFF and the W-NLFFF 
method. We show the slightly smoothed (3-point median) evolutionary
curves, from which the energy decrease 
$\Delta E_{free}=E_{free}(t_2)-E_{free}(t_1)$ is measured between the
maximum energy before, and the minimum energy after the flare peak time,
allowing for a time margin of $\pm 0.5$ hrs, i.e.,  
$t_{start}-0.5 < t_1 < t_{peak}$ and $t_{peak} < t_2 < t_{end}+0.5$. 
The free energy $E_{free}(t)$ agrees well between the VCA-NLFFF
and the W-NLFFF code, as mentioned before (Fig.~12d), but the energy
decreases during flares are much more pronounced with the VCA-NLFFF
code than with the W-NLFFF code, which may indicate that the W-NLFFF
code suffers from over-smoothing in the preprocessing procedure.

In Fig.~15 we compile the decreases of the free energy $-\Delta E_{free}$
during flares as a function of the free energy $E_{free}$ before the
flare and find that in the average 21\% of the free energy is dissipated
during flares according to the VCA-NLFFF code, or 11\% according to the
W-NLFFF code. This is similar to the earlier study (Fig.~11
in Aschwanden, Sun, and Liu 2014b). Thus there is a discrepancy of a
factor of $\approx 2$ between the VCA-NLFFF and the W-NLFFF code.

\subsection{	Performance Test : X-Class Flares 	}

In Fig.~16a and 16b we present the results of the free energy evolution
$E_{free}(t)$ for 11 X-class flare events, which includes all X-class
events that occurred in the first 3 years of the SDO mission
(Aschwanden, Xu, and Jing 2014a). For each flare we show the GOES
light curve, the GOES time derivative (which is a proxy for the
hard X-ray emission according to the Neupert effect), and the
evolution of the free energy according to both the VCA-NLFFF and
the W-NLFFF codes. Since the X-class flares are the most energetic
events, we expect that they exhibit most clearly a step function from
a high preflare value of the free energy to a lower postflare level
after the flare. Such a step function is most clearly seen with the
W-NLFFF code for flare events \#12, \#66, \#67, and \#147 (red diamonds
in Figs.~16 and 17), and with the
VCA-NLFFF code for the events \#67 and \#384 (blue curve in
Figs.~16 and 17). The flare event \#67
exhibits the best agreement between the W-NLFFF and VCA-NLFFF code,
displaying not only a large step function, but also good agreement
between the levels of free energy before and after the flare.
No significant energy decrease during the flare time interval is
detected for event \#148 with the W-NLFFF code, or for event \#147
with the VCA-NLFFF code. Generally, the VCA-NLFFF code reveals a
larger step of free energy decrease than the W-NLFFF code (Fig.~15). 
Thus we can conclude from the performance tests of the
VCA-NLFFF code: (1) A significant decrease in the free energy during
X-class flares is detected in 10 out of the 11 cases; (2) the
maximum of the energy decrease of free energy occurs within the
impulsive flare phase (when hard X-rays or the GOES time derivative
culminate), and (3) the energy decreases detected with the VCA-NLFFF 
code are a factor of $\approx 2$ larger than detected with the W-NLFFF code.

In Figs.~18 and 19 we show the results of the VCA-NLFFF solution
for the flare events \#67 (X1.8 flare on 2011 Sept 7),
\#384 (X1.2 flare on 2014 Jan 7), and
\#592 (X1.9 flare on 2014 Mar 29). 
We show two representations of the magnetic field solution, one by 
selecting field lines that intersect with the midpoints of the automatically
traced loops (Fig.~18), and the other one by a regular grid of field lines 
with footpoint magnetic field strengths of $B > 100$ G (Fig.~19).
Since these events show the most consistent evolution of the
free energy decrease with both the VCA-NLFFF and the W-NLFFF code, they
should convey most clearly the topology change from a helically twisted
nonpotential field before the flare to a relaxed near-potential field
after the flare. Fig~18a or 18b shows the NLFFF solution at 22:02 UT, just
when the free energy reaches the highest value 
($E_{free}=165 \times 10^{30}$ erg) at the start of the flare,
which indeed reveals a highly twisted field around the leading sunspot.
Fig.~18b shows the NLFFF solution at 23:08 UT, just after the impulsive
flare phase when the free energy drops to the lowest value
($E_{free}=14 \times 10^{30}$ erg), which indeed exhibits an untwisted, 
open field above the sunspot, while a postflare arcade grows in the 
eastern part of the sunspot, where obviously most of the magnetic 
reconnection process during the flare took place. The open-field
configuration above the sunspot is a consequence of
an erupting CME. This example shows an unambiguous magnetic energy
decrease by $\approx 90\%$ of the free energy
($\Delta E_{free} =-151 \times 10^{30}$ erg), which is dissipated
during a magnetic reconnection process and lauch of a CME. 

Also the second case shown in Fig.~18 and 19 (flare \#384, X1.2 class,
2014 Jan 7), exhibits an untwisting of the magnetic field above
the sunspot during the flare, most clearly seen as helical twist
in clock-wise direction (Fig.~19c) that de-rotates clock-wise
to an almost radial potential field after the flare (Fig.~19d). The
detection of loop structures appears to be spotty in Fig.~18c and 18d,
but the model picks up sufficient field directions around the leading
sunspot to measure the untwisting of the sunspot field and the associated
magnetic energy decrease.

\subsection{	Performance Test : AIA versus IRIS	 	}

The first X-class flare observed with IRIS occurred on 2014 Mar 29,
17:40 UT, which has already been modeled with a previous version of
the VCA-NLFFF code, showing a similar amount of energy decrease
in coronal data from AIA (94, 131, 171, 193, 211, 335 \ang ),
chromospheric data from AIA (304, 1600 \ang ), and in chromospheric
data from IRIS (1400, 2796 \ang ). This result delivered the 
first evidence that both coronal as well as chromospheric features
(loops or fibrils) can be used to constrain a nonpotential magnetic
field model of an active region or flare (Aschwanden 2015). 
However, the early
version of the VCA-NLFFF code was not very sensitive to the faint
coronal and chromospheric features during the preflare phase, so 
that substantially less free energy was detected during the preflare
phase than is found with the improved VCA-NLFFF code. This lack
of coronal and chromospheric structures in the preflare phase was
interpreted in terms of a coronal illumination effect, conveyed
by filling of coronal loops with the chromospheric evaporation
process. However, since the new VCA-NLFFF code is sufficiently
sensitive to detect nonpotential structures in the preflare phase,
an explanation in terms of a coronal illumination effect is not 
needed anymore.

In Fig.~18 and 19 we show new results of the evolution of the free energy
$E_{free}(t)$ for the 2014 March 29 flare, independently modeled
with the VCA-NLFFF code in three different wavelength domains and
with two independent instruments (AIA and IRIS). The representation
of the results in Fig.~20 can be compared with Fig.~3 in 
Aschwanden (2015). We detect about the same free energy at the
flare peak, $E_{free}(t_{peak}) \approx 40 \times 10^{30}$ erg
versus  $E_{free}(t_{peak})=(45 \pm 2) \times 10^{30}$ erg earlier
(Fig.~3 in Aschwanden 2015).
We find about the same decrease in the free energy, 
flare peak, $E_{free}(t_{peak} \approx 30 \times 10^{30}$ erg
versus  $E_{free}(t_{peak}=(29 \pm 3) \times 10^{30}$ erg earlier
(Fig.~3 in Aschwanden 2015).
However, what is much better than in the previous study is the
preflare level of the free energy, being significantly higher than
the postflare level, which is expected in the simplest scenario of
magnetic energy dissipation. Therefore, the present VCA-NLFFF can
be considered to be superior compared to earlier (less
sensitive) versions. The change in the magnetic configuration
during the flare shows an untwisting in counter clock-wise
direction (Figs.~19e and 19f), while the sigmoidal bundle of
field lines in the North-East sector of the sunspot 
(Fig.~19e) evolves into a near-potential dipolar postflare
loop arcade (see yellow loop tracings in Fig.~18f).

\subsection{	Performance Test : Parametric Study 	 	}

Finally we perform a parametric study in order to investigate
the robustness and sensitivity of the VCA-NLFFF solutions 
to the control parameters of the numerical code. A list of
24 control parameters is given in Table 1, which includes four
groups, one specifying the data selection (instrument, spatial
resolution, wavelength, and field-of-view choice), a second one
with 4 control parameters enables the potential field
deconvolution, a third one with 9 control parameters 
enables the automated loop tracing, and a fourth one with 7
control parameters enables the forward-fitting. In order to
assess the robustness of the VCA-NLFFF code, we investigate
how a variation of the control parameters affects the
results of the evolution of the free energy $E_{free}(t)$
(Figs.~21, 22, 23). Specifically we show how the time resolution
of the data or cadence (Fig.~21), the wavelength choice (Fig.~22),
and the variation of 12 tunable control parameters changes the
time-dependent free energy values (Fig.~23) of the X-flare
observed on 2014 March 29 with both AIA and IRIS, of which 
various data are shown in Figs.~3, 4, 9, 10, 18e, 18f, 19e, 
19f, and 20. 

\medskip

\underbar{(1) Cadence:} Varying the time resolution or cadence 
from $t_{cad}=6$ min to 3, 2, and 1 min (Figs.~21, 23a) we 
find that the smoothed time profile $E_{free}(t)$ (using the median 
values from time intervals that are equivalent to the cadence) are
invariant and thus yield exactly the same free energy decreases
associated with the flare at any value of the time resolution.
This is not trivial, because there is a factor of 6 times different
amount of information that is used between a cadence of 6 min and 1 min.
If the free energies determined with VCA-NLFFF code are entirely
due to noise, the energy drop between the preflare and postflare time
interval would vary arbitrarily, rather than being invariant.
Investigating the evolution of the free energy at a cadence of 1 min,
however, appears to reveal some coherent quasi-periodic fluctuations
with an approximate period of $P \approx 3$ min (Fig.~21 bottom left),
which could be associated with the helioseismic global p-mode 
oscillations (a property that will be examined elsewhere).
In Fig.~21 we show also the time evolution of the number of
detected and fitted loops (Fig.~21, middle column), which seems 
to be roughly constant during this flare and is not correlated
with the step-like (dissipative) decrease in free energy. We
show also the time evolution of the misalignment angle
$\mu_2(t) \approx 10^\circ$ (Fig.~21, right column), which is 
essentially constant and is not affected by the flare evolution,
although the flare area coverage of automatically traced loops
varies substantially during the flare (Fig.~18e and 18f). 

\underbar{(2) Wavelengths:} The choice of wavelengths could
crucially affect the accuracy of the inferred free energy,
because each wavelength covers only a limited amount of the
flare temperature range, and some wavelengths are only sensitive
to chromospheric temperatures. In Fig.~22 we perform the experiment
to determine the evolution of the free energy $E_{free}(t)$ for
each of the 8 AIA and 3 IRIS wavelengths separately. Interestingly,
the decrease in the free energy is detected in almost all
wavelengths independently, which provides a strong argument
for the robustness of the code. The largest energy decrease
($\Delta E_{free}=-(37 \pm 10) \times 10^{30}$ erg is detected in the
AIA 211 \ang\ wavelengths, while the smallest amount is found with
IRIS 2832 \ang , which is a chromospheric line. Combining two 
wavelengths pair-wise together, we find a consistent energy decrease in each 
wavelength pair (Fig.~23b). Consequently, our strategy is to
combine all coronal AIA wavelengths (94, 131, 171, 193, 211, 335 \ang )
as a default, because the joint wavelength response complements
hot and cool temperatures in flares and active region areas.

\underbar{(3) Minimum loop lengths:} Since the number of loops
above some minimum length value $l_{min}$ falls off like a powerlaw
distribution function, i.e., $N(>L) \propto L^{-2}$ 
(Aschwanden et al.~2008), the number of loops that constrain
a NLFFF solution increases drastically towards smaller values
on one side, while the ambiguity of true loops and loop-unrelated
curved features increases towards smaller values on the other hand.
The test shown n Fig.~23c exemplifies that a value of $l_{min}\approx
4$ or 5 is the optimum, while higher values of $l_{min}=6$ or 7
lead to slight underestimates of the free energy.

\underbar{4) Misalignment angle limit:} Allowing a large range of
2D misaligned angles $\mu_0 < 45^\circ$ clearly hampers the convergence
of the VCA-NLFFF code, which implies that data noise dominates the
fit and thus reduces the signal of nonpotential field energies and
free energies. The test shown in Fig.~23d demonstrates that the
free energy increases systematically by reducing the misalignment
limit from $\mu_0 < 40^\circ$ to $\mu_0 < 10^\circ$. However, if we
push the limit to small values, the number of fitable features 
decreases, which has a diminuishing effect on the accuracy of a
nonpotential field solution. Therefore we choose
a compromise of $\mu_0 < 20^\circ$.

\underbar{5) Minimum number of iterations:} The number of iterations
in the forward-fitting algorithm dictates how quickly misaligned
features are eliminated, say for an acceptable limit of 
$\mu_0 < 20^\circ$. Reducing the iterations thus changes the
observational constraints of fitable loops faster than the code
can converge, and thus can inhibit convergence of the code.
The test shown in Fig.~23e clearly demonstrates that the free
energy is only fully retrieved for a larger number of iterations,
say $n_{nitmin} \gapprox 40$.

\underbar{6) Spatial smoothing of EUV image:} The test shown in
Fig.~23f indicates that smoothing of the EUV loop image with
$n_{smo}=1, 2$, or 3 reduces the estimate of the free energy
because faint and thin loop structures are eliminated, which
reduces the observational constraints and accuracy of the 
forward-fitting code.

\underbar{7) Number of magnetic sources:} We expect that 
increasing the number of magnetic sources increases the accuracy
of the free energy. The test shown in Fig.~23g, however shows the
opposite, probably because too many small magnetic sources
counter-balance the opposite polarities of closely-spaced pairs
of magnetic sources, and thus diminuishes the total nonpotential
or free energy. We thus choose a relatively small number of
$n_{mag}=30$ magnetic sources, which also speeds up the
computation time of a VCA-NLFFF solution.

\underbar{8) Flux and flux filter thresholds:} The tests shown
in Fig.~23h and 23i indicate that the threshold of detecting
loop structures is not critical, as long as we detect a sufficient
number of structures. We choose therefore the lowest threshold
values of $q_{thresh,1}=0$ and $q_{thresh,2}=0$. 

\underbar{9) Magnetic proximity:} In a previous version of the
VCA-NLFFF code we used a distance limit $d_{prox}$ of a loop
position to the next location of a magnetic source to eliminate
``false'' loop detections. The test shown in Fig.~23l demonstrates
that large distance limits of $d_{prox}=4$ or 10 source depths
(which is about equivalent to the apparent full width of a magnetic 
source at the solar surface) does not effect the accuracy of the
free energy estimate, while very short limits of $d_{prox}=1$ or
2 eliminate too many relevant structures and thus leads to
underestimates of the free energy.

\underbar{10) Loop curvature radius limit:} The lower limit of the
curvature radius of automatically detected structures is important
to exclude ``false'' curvi-linear structures that occur through
coagulation of random structures. The test in Fig.~23m with
curvature radii limits of $r_{min}=4, 6, 8,$ and 10 pixels demonstrates
an invariant free energy profile, and thus no sensitivity of the
free energy to the curvature radius limit.

\underbar{11) Field-of-view:} If the field-of-view is too small,
the free energy is retrieved only partially. The test shown in
Fig.~23n indicates indeed a steady increase of the free energy
from $FOV=0.08$ to 0.14 solar radii. On the other side, an upper
limit of the FOV is given by the distance to the next neighboured
active region. 

\medskip
In summary, our parametric tests are performed on 12 parameters,
each one with 4 variations, for a time profile of the free energy
with 13 time steps, yielding a total of 624 VCA-NLFFF solutions.
The test results shown in Fig.23 show us at one glance that all
of the 48 time profiles exhibit clearly the step-wise decrease
of the free energy during a flare, corroborating the robustness
of our VCA-NLFFF code. Even the magnitude of the energy decrease
agrees within $\approx 20\%$ for each parameter combination. 

An estimate of the uncertainty in the determination of the free 
energy can be made by the number of loops $n_{loop}$ that 
constrain a solution, which is expected to be according to
Poisson statistics, 
\begin{equation}
	\sigma_{E,free} = {E_{free} \over \sqrt{n_{loop}}} \ .
\end{equation}
In other words, if only one single loop is used in fitting the
VCA-NLFFF magnetic field model, we have an error of 100\%, while
the error drops down to \lapprox 10\% for $n_{loop}\approx 100-200$, 
which is a typical value (Fig.~21, middle column). However, this
error estimate is a lower limit, applicable to the ideal case 
when a sufficient number of loops is available in the locations of
strong magnetic fields. If there is an avoidance of loops
near sunspots, no significant amount of free energy is detected,
leading to a much larger systematic error than the statistical
error due to Poisson statistics of the number of loops. 

\section{	    DISCUSSION 			}

The main purpose of the VCA-NLFFF code is the measurement of the
coronal magnetic field and its time evolution in active regions
and during flares, based on our vertical-current approximation
model. The following discussion mostly focuses on the example 
of active region NOAA 11158 and its famous X2.2 flare in the
context of previous studies.

\subsection{	Magnetic Field Changes in Active Region NOAA 11158 	}

A most studied example is the evolution of {\bf active region NOAA 11158}, 
which is the subject of a number of recent papers, focussing on the 
magnetic evolution of this active region (Sun et al.~2012a, 2012b, 2015; 
Liu and Schuck 2012; Jing et al.~2012; 
Vemareddy et al.~2012a, 2012b, 2013; 2015;
Petrie 2012; Tziotziou et al.~2013; Inoue et al.~2013, 2014; 
Liu et al.~2013; Jiang and Feng 2013; 
Dalmasse et al.~2013; Aschwanden et al.~2014b; Song et al.~2013;
Zhao et al.~2014; Tarr et al.~2013; Toriumi et al.~2014; 
Sorriso-Valvo et al.~2015; Jain et al.~2015; Guerra et al.~2015;
Kazachenko et al.~2015; Li and Liu 2015; Chintzoglou and Zhang 2013;
Zhang et al.~2014; Gary et al.~2014),
or specifically on the {\bf X2.2 flare that occurred on 2011 February 15}
in this active region (Schrijver et al.~2011; Wang et al.~2012;
Aschwanden et al.~2013b; Jiang et al.~2012; Gosain 2012;
Maurya et al.~2012; Petrie 2013; Alvarado-Gomez et al.~2012;
Young et al.~2013;
Malanushenko et al.~2014; Wang et al.~2014; Beauregard et al.~2012;
Inoue et al.~2015; Wang et al.~2013; Shen et al.~2013; Jing et al.~2015;
Raja et al. 2014), or on the {\bf M6.6 flare that occurred on 
2011 February 13} in the same active region (Liu et al.~2012, 2013; 
Toriumi et al.~2013). 

Comparing the workings of the VCA-NLFFF code with the theoretical concepts
and observational results that have been published on AR 11158, we have
to be aware that the vertical-current approximation used in the VCA-NLFFF
code implies a slow helical twisting of the sunspot-dominated
magnetic field during the energy storage phase, and sporadic
episodes of reconnection-driven untwisting during flare times. 
Therefore, this concept is equivalent to the concept of sunspot rotation 
(Jiang et al.~2012; Vemareddy et al.~2012a, 2012b, 2013, 2015; 
Li and Liu 2015). Sunspot rotation forms sigmoid structures
naturally, as it is observed in AR 11158 (Schrijver et al.~2011; 
Sun et al.~2012a; Jiang et al.~2012; Young et al.~2013; 
Jiang and Feng 2013; Aschwanden et al.~2014b). 
Strongly twisted magnetic field lines ranging from
half-turn to one-turn twists were found to build up just before
the M6.6 and X2.2 flare and disappear after that, which is believed
to be a key process in the production of flares (Inoue et al.~2013;
Liu et al.~2013). The vortex in the source field suggests that the
sunspot rotation leads to an increase of the non-potentiality
(Song et al.~2013).

In addition, the energy storage in active region
11158 is also supplied by fast flux emergence and strong shearing
motion that led to a quadrupolar ($\delta$-type) sunspot complex 
(Sun et al.~2012a, 2012b; Jiang et al.~2012; Liu and Schuck 2012;
Toriumi et al.~2014).
Both the upward propagation of the magnetic and current helicities
synchronous with magnetic flux emergence contribute to the gradual
energy build up for the X2.2 flare (Jing et al.~2012; Tziotziou et al.~2013).
The amount of nonpotential field energies stored in AR 11158 before
the X2.2 GOES-class flare was calculated to 
$E_{np}=2.6 \times 10^{32}$ erg (Sun et al.~2012a), 
$E_{np}=5.6 \times 10^{32}$ erg (Tarr et al.~2013), 
$E_{np}=1.0 \times 10^{32}$ erg (COR-NLFFF; Aschwanden et al.~2014b), 
$E_{np}=2.6 \times 10^{32}$ erg (W-NLFFF; Aschwanden et al.~2014b), 
$E_{np}=10.6 \times 10^{32}$ erg (Kazachenko et al.~2015), 
$E_{np}=2.8 \times 10^{32}$ erg (W-NLFFF: Fig.~16), 
$E_{np}=1.8 \times 10^{32}$ erg (VCA-NLFFF: Fig.~16), 
which vary within an order of magnitude. 
The amount of dissipated energy during the X2.2 flare was calculated to
$\Delta E_{free}=-1.7 \times 10^{32}$ erg (Tarr et al.~2013), 
$\Delta E_{free}=-1.0 \times 10^{32}$ erg (Malanushenko et al.~2014), 
$\Delta E_{free}=-0.3 \times 10^{32}$ erg (Sun et al.~2015), 
$\Delta E_{free}=-0.6 \times 10^{32}$ erg (COR-NLFFF; Aschwanden et al.~2014b), 
$\Delta E_{free}=-0.6 \times 10^{32}$ erg (W-NLFFF; Aschwanden et al.~2014b), 
$\Delta E_{free}=-1.0 \times 10^{32}$ erg (W-NLFFF: Fig.~16), 
$\Delta E_{free}=-0.3 \times 10^{32}$ erg (VCA-NLFFF: Fig.~16),
which agree within an factor of $\approx 6$. 

Untwisting of helical fields (as assumed in the VCA-NLFFF model)
may not be the full explanation of the magnetic field evolution
during flares. Since the VCA concept assumes twisting around
vertically oriented axes, untwisting would reduce the azimuthal
field component $B_{\varphi}$, which corresponds to a reduction
of the horizontal field components $B_x$ and $B_y$ (for a location
near disk center), as it is the case for the X2.2 flare on 2011 February 15
in NOAA 11158. In contrast, however, the response of the photospheric
field to the flare was found to become more horizontal after eruption,
as expected from the tether-cutting reconnection model 
(Wang et al.~2012; Liu et al.~2012, 2013; Inoue et al.~2014, 2015), 
or from the {\sl coronal implosion scenario} (Gosain 2012; Wang et al.~2014). 
On the other side, reduction of
magnetic twist was explained by a large, abrupt, downward vertical
Lorentz-force change (Petrie 2012, 2013). 

In summary, twisting and untwisting of magnetic field lines, the
basic concept of our vertical-current approximation model in describing 
the magnetic evolution, plays a leading role in many of the theoretical
and observationally inferred flare models of the X2.2 flare in
active region NOAA 11158, and thus justifies the application of
the VCA-NLFFF code to flares in general, although the underlying
analytical formulation encapsulates only on particular family of
solutions among all possible nonlinear force-free field models. 

\subsection{    The Pro's and Con's of NLFFF Codes      }

After we performed non-potential magnetic field modeling with
chromospheric and coronal data we can assess the feasibility
of NLFFF modeling in a new light, which we contrast here 
between the W-NLFFF code and our VCA-NLFFF code.
Standard W-NLFFF codes have the following features:
(1) They are using photospheric vector magnetograph data
(where the transverse field components have a much larger
degree of noise than the line-of-sight component),
(2) use the assumption of a force-free photosphere,
(3) have no mechanism to fit the magnetic field solution to
observed chromospheric or coronal features,
(4) use a preprocessing method to make the photospheric
boundary condition more force-free,
(5) are designed to converge to a divergence-free and force-free solution
as accurately as possible,
(6) map the sphericity of the Sun onto a plane-parallel boundary, and
(7) are computationally expensive (with computation times
in the order of hours or days).
In contrast, the VCA-NLFFF code can be characterized
in the following way:
(1) It uses only the line-of-sight magnetic field component
and avoids the noisy transverse components;
(2) do not use the assumption of a force-free photosphere;
(3) has the capability to minimize the difference between
the theoretical magnetic field model and observed curvi-linear
features in coronal and/or chromospheric images;
(4) does not modify the photospheric boundary condition with
a preprocessing method;
(5) fulfills the divergence-free and force-free conditions
with second-order accuracy;
(6) takes the sphericity of the Sun fully into account, and
(7) is computationally relatively fast (with computation times
in the order of minutes).
Comparing these seven characteristics, it appears that the
VCA-NLFFF code has a superior design in six out of the
seven criteria. The only design where the W-NLFFF code
has a superior performance is item (5), since they are 
designed to optimize divergence-freeness, force-freeness,
and constancy of the $\alpha$-parameter along each field line,
while the VCA-NLFFF code uses an approximate analytical solution
that is accurate to second order in $\alpha$ only. However,
the reduced accuracy of the VCA-NLFFF code outweighs the
biggest short-coming of the W-NLFFF code, i.e., that it is unable
to optimize the match between model and observations.
The two types of NLFFF codes are truly complementary and one
might use the VCA-NLFFF code as a first initial guess that can
quickly be calculated before a more accurate NLFFF solution
with the W-NLFFF code is attempted. Perhaps it could provide
improved boundary conditions for the W-NLFFF code, exploiting
both chromospheric and coronal constraints.

\subsection{	Open Problems: Questions and Answers		}  					
Although this benchmark test of the VCA-NLFFF code presented here
is the most comprehensive performance test carried out to-date, 
which demonstrates encouraging results, there are still a number 
of open problems that should be pursued in future studies.
These open problems concern fundamental limitations of the 
VCA-NLFFF method, such as: (1) The suitability of the magnetic
model; (2) the fidelity of automated feature tracking; (3) the
sensitivity to nonpotential field components; (4) the
ambiguity of nonpotential field solutions; and (5) the
usefulness in practical applications, in particular for
forecasting the magnetic evolution of active regions and flares.
We discuss these five open issues briefly.

(1) The magnetic field model used in the VCA-NLFFF model is based
on buried magnetic charges and helically twisted field lines 
(or flux tubes) with an azimuthal field component $B_{\varphi}$ 
that is 
caused by vertical currents at the photospheric boundaries where
the unipolar magnetic charges are buried. There are alternative
nonlinear force-free magnetic field models, such as that of a 
sheared arcade (e.g., see textbooks of Priest 1982; Sturrock 1994;
Aschwanden 2004), for which an analytical approximation could be
derived and fitted to observed loop geometries. In principle,
the suitability of such alternative models could be tested by
building forward-fitting codes that are equivalent to the
VCA-NLFFF code, and by comparing the best-fit misalignment
angles between the models and data. More generally, a force-free
minimization code such as the W-NLFFF code could be designed
that fits coronal loops without using the transverse components
at the photospheric boundary. Such a concept was pioneered by
Malanushenko et al.(~2009, 2011, 2012, 2014), but the major
drawbacks of that code using a quasi Grad-Rubin method is the
lack of an automated loop-tracing capability and the prohibitively
long computation times.

(2) The fidelity of automated feature tracking depends on the
complexity and noise level of the used EUV and UV images, and whether there 
are a sufficient number of curvi-linear features that can be detected in
the data. The usage of AIA images has demonstrated that the
biggest challenge is the elimination of ``false'' loop tracings,
which occur due to insufficient spatial resolution, over-crossing 
structures, moss structures (Berger et al.~1999; De Pontieu et al.~1999), 
data noise, and instrumental effects (image edges, vignetting, 
pixel bleeding, saturation, entrance filter mask, etc.).
The usage of IRIS and IBIS images, which have 4-6 times higher
spatial resolution than AIA, provide a much larger amount of resolved
curvi-linear features that are helpful in the reconstruction of
the magnetic field, even when they display chromospheric features
(fibrils) rather than coronal structures (active region loops and
post-flare loops) (Aschwanden, Reardon, and Jess 2016). 
Future versions of the VCA-NLFFF code may combine
the most suitable features in coronal and chromospheric wavelengths.

(3) The sensitivity to nonpotential field components depends
on the availabiliy of detectable curvi-linear features in the
penumbral regions of sunspots, where the largest magnetic field
strengths occur outside the umbra, while the umbra is generally
void of loops and fibrils. The detectability of penumbral
structures works best for heliographic locations near the Sun
center, while regions faw away from the Sun center generally cause
confusion between loops in the foreground and moss structures
in the background plages. The accuracy of the total free energy
of an active region or flare is dominated by the magnetic field
structures near the leading or dominant sunspot, because of the
highly nonlinear $B^2$-dependence of the free energy on the
twisted field. Thus the sensitivity of the VCA-NLFFF code to
the free energy is determined by the detectability of structures 
near the dominant sunspot, in contrast to W-NLFFF codes, where
the sensitivity of the free energy is limited by the noise and
uncertainty of the transverse field components $B_x$ and $B_z$
in the dominant sunspot region. 

(4) The VCA-NLFFF code has typically $n_{mag}=20-100$ magnetic
source components, and thus the numerical convergence of the
VCA-NLFFF code towards a best-fit solution is in principle
not unique, given the uncertainties of automatically traced
curvi-linear features. However, we have demonstrated in the
parametric study (Section 5.6) that our results are extremely 
robust when
the various control parameters of the VCA-NLFFF code are varied,
which implies a stable convergence to the same solution. 
What we can say about the convergence ambiguity is that the
solution is dominated by the strongest magnetic sources, 
which implies near-unambiguity for the strongest sources
(such as the dominant sunspot), while the degree of ambiguity
increases progressively for weaker magnetic sources. Since
we are mostly interested in the total (volume-integrated) free
energy, which is also dominated by the strongest sources due
to the $B^2$-dependence, the value of the free energy 
$E_{free}(t)$ is nearly unambiguous, at any instant of time.

(5) What is the usefulness of the VCA-NLFFF code for practical 
applications. The main product of the VCA-NLFFF code is the
free energy and its time evolution, $E_{free}(t)$, which includes
also estimates of the dissipated magnetic energy during flares,
i.e., $E_{diss}=E_{free}(t_{end})-E_{free}(t_{start})$.
In principle, other W-NLFFF codes can provide the same information,
but their uncertainty is not known, because the inferred values
may be affected by the violation of the force-freeness in the
photosphere and lower chromosphere, the mismatch between the
NLFFF model magnetic field and the observed loops, and the
smoothing of the preprocessing technique. Ideally, if the
evolution of the free energy can be obtained with either method,
this parameter is one of the most relevant quantities for
flare forecasting. A machine-learning algorithm called {\sl support
vector machine (SVM)} has been applied to an HMI data base of
2071 active regions and it was found that the total photospheric
magnetic free energy density is the third-best predictor for
flares (out of 25 tested quantities), besides the total unsigned 
current helicity and the total magnitude of the Lorentz force 
(Bobra and Couvidat 2015). 

\section{		CONCLUSIONS				}

In this study we present an updated description and performance
tests of the vertical current approximation nonlinear force-free
field (VCA-NLFFF) code, which is designed to calculate the free
energy $E_{free}(t)$ and its time evolution in active regions and 
flares, after it has been continuously developed over the last five 
years. This code provides a complementary and alternative method
to existing traditional NLFFF codes, such as the W-NLFFF code
(Wiegelmann 2004). The VCA-NLFFF method requires the input of
a line-of-sight magnetogram and automated curvi-linear tracing
of EUV or UV images in an arbitrary number of wavelengths and
instruments. In contrast, the W-NLFFF code requires the input
of 3D magnetic field vectors in the photospheric boundary, but
has no capability to match observed features in the corona or 
chromosphere. The performance tests presented here include 
comparisons of the potential, non-potential, free energy, and 
flare-dissipated magnetic energy between the VCA-NLFFF and the
W-NLFFF code. We summarize the major conclusions:

\begin{enumerate}

\item{The chief advantages of the VCA-NLFFF code over the W-NLFFF
	code are the circumvention of the unrealistic assumption
	of a force-free photosphere in the magnetic field 
	extrapolation method, the capability to minimize
	the misalignment angles between observed coronal
	loops (or chromospheric fibril structures) and 
	theoretical model field lines, as well as computational 
	speed.}

\item{Comparing 600 W-NLFFF solutions from active region NOAA 11158
	and 119 W-NLFFF solutions from 11 X-class flares with 
	VCA-NLFFF solutions we find agreement in the potential,
	non-potential, and free energy within a factor of 
	$\approx 1.2$, which compares favorably with respect to
	the range of free energies that have been obtained with 
	other NLFFF codes, scattering by about an order of 
	magnitude for published values of the X2.2 flare 
	in NOAA 11158.}

\item{The time evolution of the free energy $E_{free}(t)$ in
	11 X-class flares modeled with both NLFFF codes
	yields a significant decrease of the free energy
	during the flare time interval in 10 out of the 11 cases, 
	but the the energy amount determined with the W-NLFFF
	code is statistically a factor of 2 lower, probably
	because of over-smoothing in the preprocessing 
	technique. In addition we tested the VCA-NLFFF code
	for 36 C, M, and X-class flares in AR 11158 and
	detected a significant energy decrease in most cases.}

\item{The amount of magnetic energy decrease during flares
	agrees within a factor of $\approx 2-3$ between
	the two NLFFF codes. We suspect that the VCA-NLFFF
	code fails to detect the full amount of magnetic
	energy decreases in cases with insufficient loop
	coverage in penumbral regions. In contrast, the
	W-NLFFF code may fail to detect the full amount
	of magnetic energy decreases as a consequence of
	over-smoothing in the preprocessing procedure and
	the violation of the force-free condition in the
	photosphere.}

\item{Both the VCA-NLFFF and the W-NLFFF codes are able to
	measure the magnetic energy evolution in active regions
	and the magnetic energy dissipation in flares, 
	but each code has different systematic errors,
	and thus the two types of codes are truly complementary.
	The present absolute error in the determination
	of changes in the free energy during large flares,
	due to systematic errors, is about a factor of 2,
	based on the discrepancy between the compared two codes.}
		
\end{enumerate}

The obtained results are encouraging to justify further developments
of the VCA-NLFFF code. Future studies may focus on a deeper 
understanding of the systematic errors of various NLFFF codes,
which will narrow down the accuracy of free energies and
flare-dissipated energies. The depth of the buried magnetic charges
inferred from the VCA-NLFFF code may give us deeper insights
into the solar dynamo and local helioseismology results, an
aspect that we have not touched on here. Correlation studies of 
time series of the free energy may reveal new methods to improve flare
forecasting in real time. 

The VCA-NLFFF code is also publicly available in the
{\sl Solar SoftWare (SSW)} library encoded in the
{\sl Interactive Data Language (IDL)}, see website
{\sl http://www.lmsal.com/$\sim$aschwand/software/}. 

\bigskip
\acknowledgements
The author is indebted to helpful discussions with Bart De Pontieu,
Mark DeRosa, Anna Malanushenko, Carolus Schrijver, Alberto Sainz-Dalda, 
Ada Ortiz, Jorrit Leenarts, Kevin Reardon, and Dave Jess. 
Part of the work was supported by the NASA contracts NNG04EA00C 
of the SDO/AIA instrument and NNG09FA40C of the IRIS mission.

\clearpage



\begin{deluxetable}{lll}
\tabletypesize{\normalsize}
\tablecaption{Data selection parameters and adjustable control parameters 
of the VCA-NLFFF forward-fitting code used in this study.}
\tablewidth{0pt}
\tablehead{
\colhead{Task:}&
\colhead{Control parameter}&
\colhead{Value}}
\startdata
Data selection: &Instruments 			& HMI; AIA; IRIS; IBIS; ROSA   \\
                &Spatial pixel size             & $0.5\arcsec$; $0.6\arcsec$; 
                                                  $0.16\arcsec$; $0.1\arcsec$; $0.1\arcsec$ \\
                &Wavelengths                    & 6173; [94,131,171,193,211,304,335,1600]; \\
		&				& [1400,2796,2832]; 8542; 6563 \ang \\
                &Field-of-view                  & $FOV = 0.1,...,0.4 R_{\odot}$\\
Magnetic sources: &Number of magnetic sources   & $n_{mag} = 30$            \\
		&Width of fitted local maps     & $w_{mag}=3$ pixels \\
		&Depth range of buried charges  & $d_{mag} = 20$ pixels \\
		&Rebinned pixel size            & $\Delta x_{mag}=3$ pixels = $1.5\arcsec$ \\
Loop tracing:   &Maximum of traced structures   & $n_{struc}=1000$	       \\
		&Lowpass filter   	        & $n_{sm1} = 1$ pixel         \\
                &Highpass filter                & $n_{sm2} = n_{sm1}+2 = 3$ pixels\\
                &Minimum loop length            & $l_{min} = 5$ pixels         \\
                &Minimum loop curvature radius  & $r_{min} = 8$ pixels         \\
                &Field line step                & $\Delta s=0.002 R_{\odot}$ \\
		&Threshold positive flux        & $q_{thresh,1} = 0$           \\
		&Threshold positive filter flux & $q_{thresh,2} = 0$           \\
                &Proximity to magnetic sources  & $d_{prox}=10$ source depths  \\
Forward-Fitting:&Misalignment angle limit       & $\mu_0 = 20^\circ$           \\
                &Minimum number of iterations   & $n_{iter,min}= 40$           \\
                &Maximum number of iterations   & $n_{iter,max}= 100$          \\
		&Number loop segment positions  & $n_{seg}=9$			\\
		&Maximum altitude		& $h_{max}=0.2 R_{\odot}$ 	\\
		&$\alpha$-parameter increment   & $\Delta \alpha_0=1.0 \ R_{\odot}^{-1}$ \\
		&Isotropic current correction   & $q_{iso}=(\pi/2)^2\approx 2.5$ \\
\enddata
\end{deluxetable}


\begin{deluxetable}{rlllrllll}
\tabletypesize{\normalsize}
\tablecaption{Data sets used in performance tests of the VCA-NLFFF
code in this study. The flare number corresponds to a continuous
numbering of M- and X-class flares since start of the SDO mission.
The duration includes a time interval of $\pm 1$ (or $\pm0.5$) hrs
before and after flares.}
\tablewidth{0pt}
\tablehead{
\colhead{Flare}&
\colhead{Start time}&
\colhead{Duration}&
\colhead{Cadence}&
\colhead{Time steps}&
\colhead{GOES}&
\colhead{Heliographic}&
\colhead{NOAA}&
\colhead{Instrument}\\
\colhead{number}&
\colhead{(UT)}&
\colhead{}&
\colhead{NLFFF,AIA}&
\colhead{NLFFF,AIA}&
\colhead{class}&
\colhead{Position}&
\colhead{AR}&
\colhead{}}
\startdata
10-15	&2011 Feb 12-17	    & 5 days  &12, 6 min & 600, 1200  & C1.0-X2.2 & S20E27-W38 & 11158 & SDO \\
	&		    &	      &		 &	      &	    &            &       &     \\
12	&2011 Feb 15, 00:44 & 2.4 hrs &12, 6 min & 12, 24     & X2.2	  & S21W12     & 11158 & SDO \\
37	&2011 Mar 09, 22:13 & 2.3 hrs &12, 6 min & 11, 23     & X1.5      & N10W11     & 11166 & SDO \\
66	&2011 Sep 06, 21:12 & 2.3 hrs &12, 6 min & 11, 23     & X2.1 	  & N16W15     & 11283 & SDO \\
67	&2011 Sep 07, 21:32 & 2.3 hrs &12, 6 min & 6,  12     & X1.8      & N16W30     & 11283 & SDO \\
147	&2012 Mar 06, 23:02 & 2.7 hrs &12, 6 min & 13, 27     & X5.4      & N18E31     & 11429 & SDO \\
148	&2012 Mar 07, 00:05 & 2.4 hrs &12, 6 min & 12, 24     & X5.4 	  & N18E29     & 11430 & SDO \\
220	&2012 Jul 12, 14:37 & 3.9 hrs &12, 6 min & 19, 39     & X1.4      & S15W05     & 11520 & SDO \\
344	&2013 Nov 05, 21:07 & 2.2 hrs &12, 6 min & 11, 22     & X3.3      & S08E42     & 11890 & SDO \\
349	&2013 Nov 08, 03:20 & 2.2 hrs &12, 6 min & 11, 22     & X1.1      & S11E11     & 11890 & SDO \\
351	&2013 Nov 10, 04:08 & 2.2 hrs &12, 6 min & 11, 22     & X1.1      & S11W17     & 11890 & SDO \\
384	&2014 Jan 07, 17:04 & 3.0 hrs &12, 6 min & 15, 30     & X1.2      & S12E02     & 11944 & SDO \\
	&		    &	      &		 &	      &           &            &       &     \\
592	&2014 Mar 29, 17:05 & 1.3 hrs &..., 6 min & ...,  13  & X1.0      & N10W33     & 12017 & SDO,IRIS \\
\enddata
\end{deluxetable}

\clearpage


\begin{figure}
\plotone{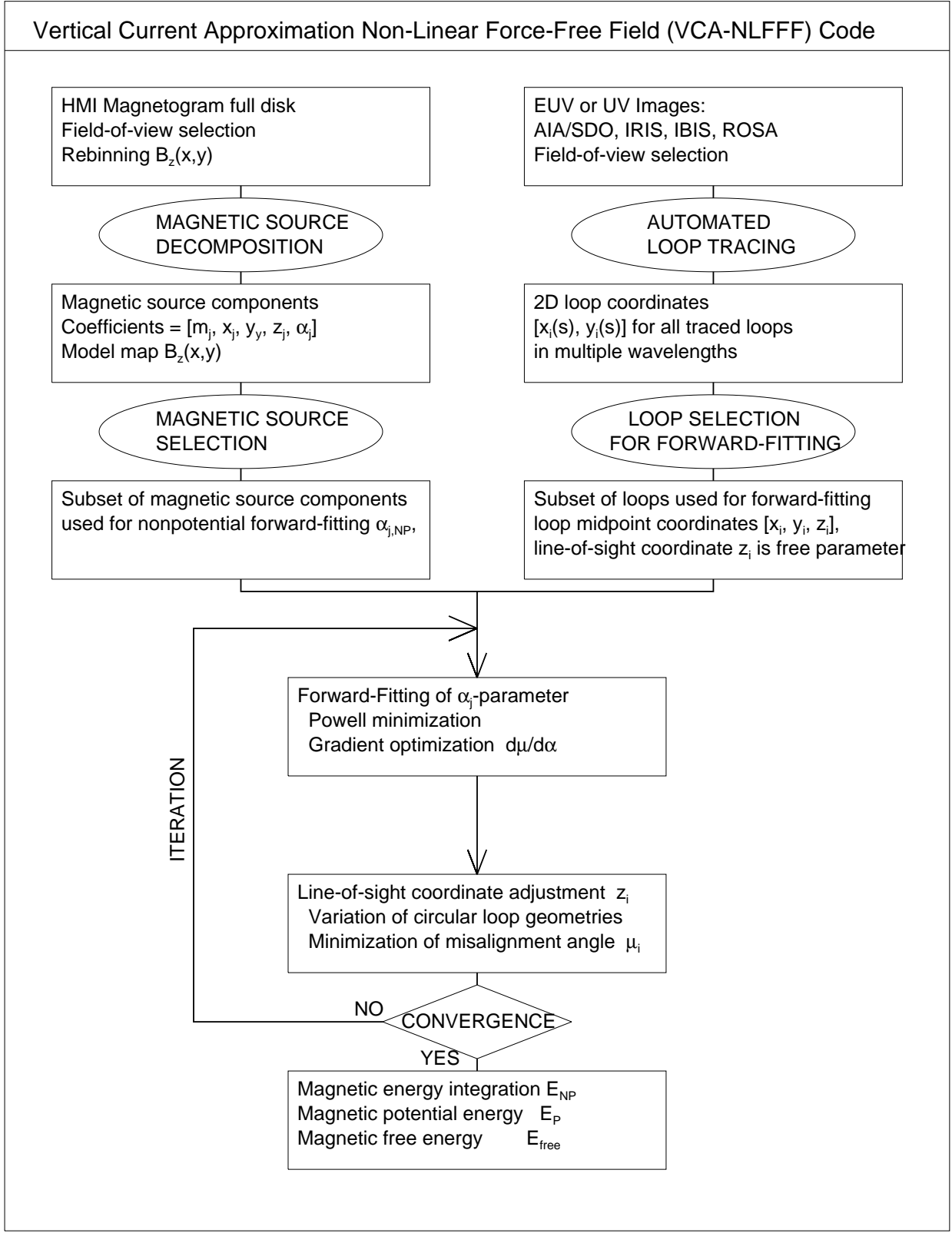}
\caption{Flow chart of the VCA-NLFFF code, which includes a decomposition
of the magnetic data (top left), automated tracing of curvi-linear structures
(top right), and forward fitting using both data sets (bottom half)
(adapted from Aschwanden et al.~2014b).}
\end{figure}
\clearpage

\begin{figure}
\plotone{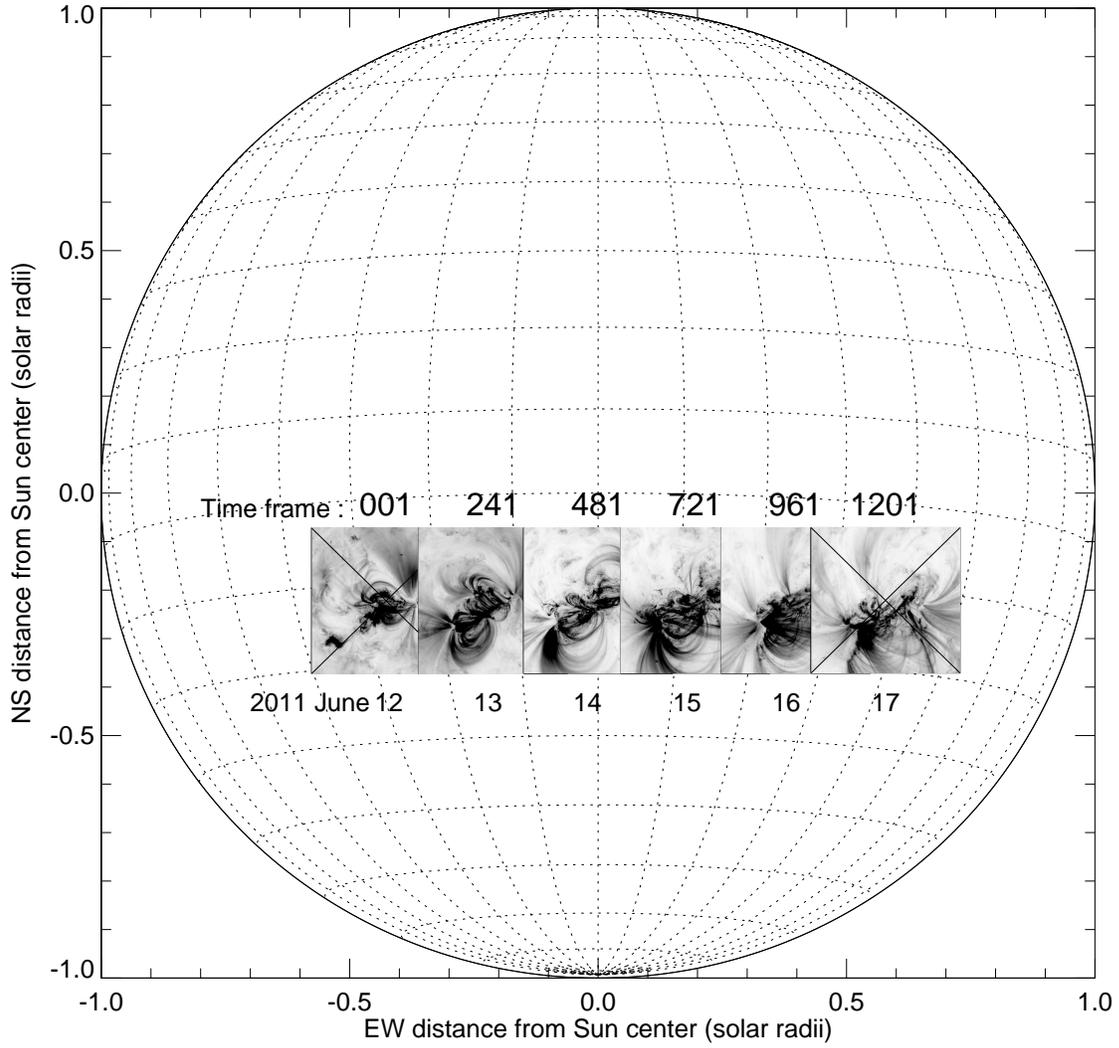}
\caption{An example of a time grid with 1200 steps is shown for AR 11158 on
2011 February 12 through 17, showing the time-dependent field-of-view
at the beginning of each day (00:00 UT). The x-axis and y-axis indicate the
cartesian coordinate system, while a grid of longitudes and latitudes with 
$10^\circ$ spacing illustrates the heliographic coordinate system. The 
heliographic position of a selected time frame refers to the center of the 
subimage (indicated with a diagonal cross in the first and last frame.}
\end{figure}
\clearpage

\begin{figure}
\plotone{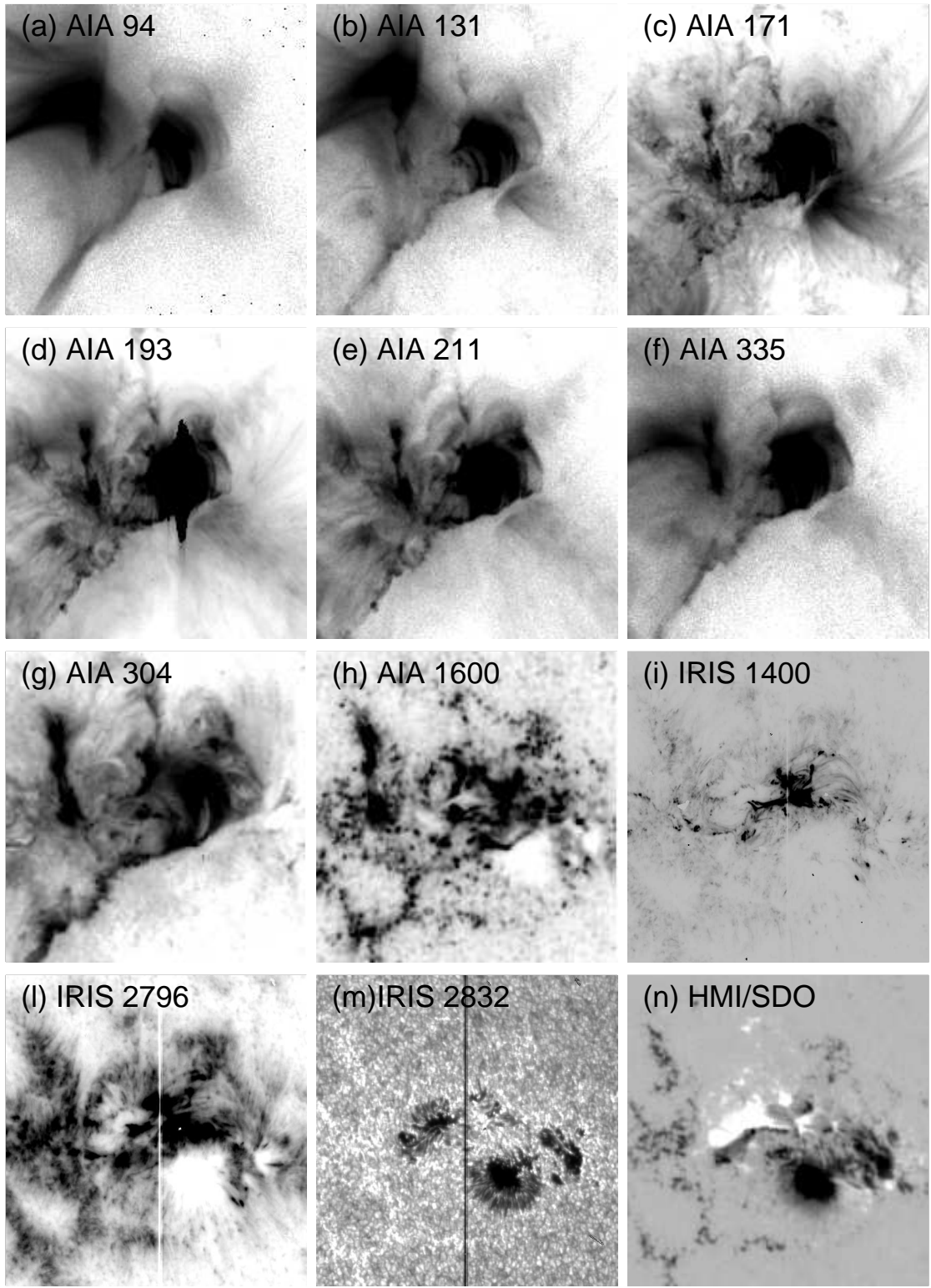}
\caption{Example of an image set that is used for modeling with the
VCA-NLFFF code for a single time step, containing subimages in 8 AIA/SDO 
wavelengths (panels a-h), 3 IRIS wavelengths (panels i-m), and the 
corresponding HMI/SDO magnetogram (panel n). This dataset 
has been observed on 2014 March 29, 18:17 UT with AIA, and at 17:35 UT with IRIS,
shown for a field-of-view of 0.1 solar radii, centered at heliographic 
position N10W33. The greyscale of the images is inverted (panels a-j), 
except for IRIS 2832 and HMI (panels m,n), on a logarithmic intensity scale 
for AIA images (panel a-g), and on a linear scale for the others (panel h-n).
The AIA images show a compact postflare loop arcade, while IRIS images reveal 
chromospheric features. CCD saturation is visible in the AIA 195 \ang\ image 
(panel d).}
\end{figure}
\clearpage

\begin{figure}
\plotone{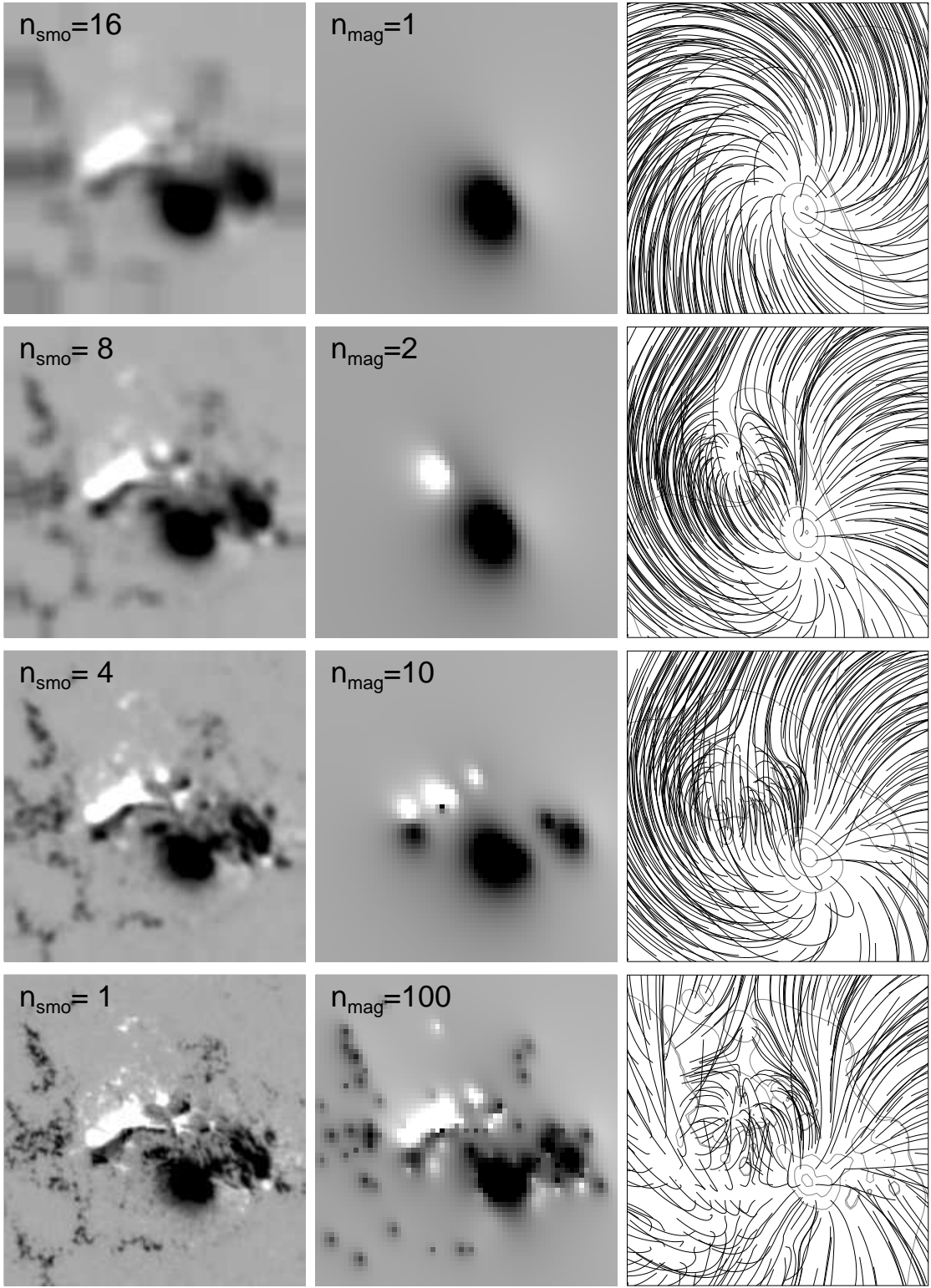}
\caption{The decomposition of a line-of-sight magnetogram $B_z(x,y)$ (bottom
left) into a finite number of $n_{mag}=1$, 2, 10, or 100 magnetic sources 
(middle column) is shown with the corresponding potential field (field lines 
in right column), overlayed on the (grey) contours of the magnetogram.
For comparison, a spatial smoothing of the magnetogram is shown also,
with smoothing boxcars of $n_{smo}=4$, 8, and 16 pixels of the original 
HMI resolution (panels in left column).}
\end{figure}
\clearpage

\begin{figure}
\plotone{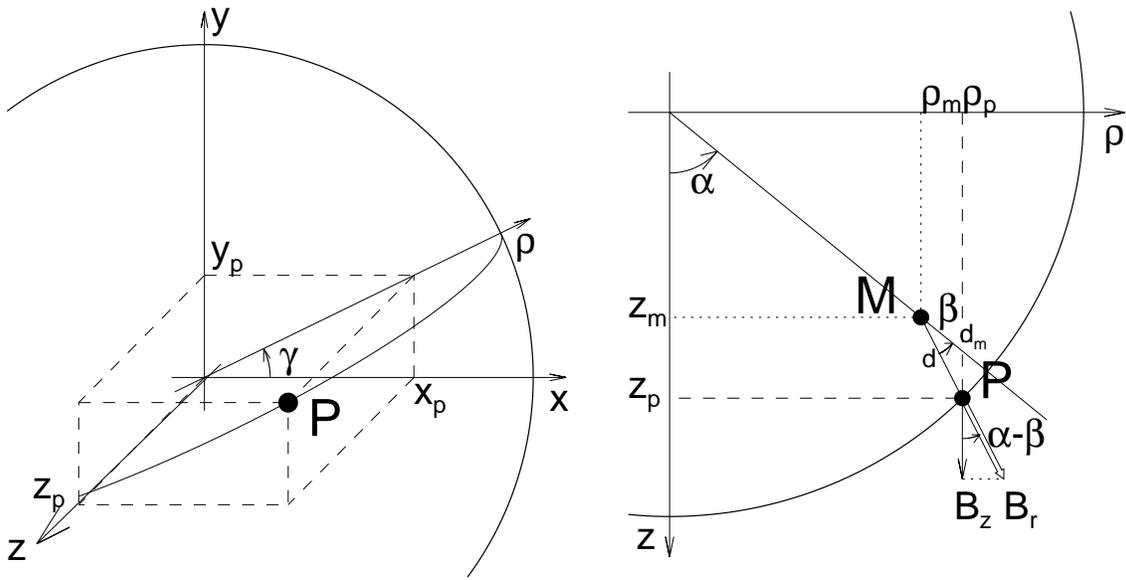}
\caption{3D geometry of a point source $P=(x_p, y_p, z_p)$
in a cartesian coordinate system is shown (left), with the $z$-axis aligned
to the line-of-sight from Earth to Sun center. The plane through the
line-of-sight axis and the point source $P$ has a position angle $\gamma$
in the plane-of-sky with respect to the $x$-axis and defines the direction
of the axis $\rho$.
The geometry of a line-of-sight magnetic field component $B_z$ is shown
in the $(z,\rho)$-plane on the right hand side. A magnetic point charge $M$
is buried at position $(z_m, \rho_m)$ and has an aspect angle $\alpha$
to the line-of-sight. The radial component $B_r$ is observed on the solar
surface at location $P$ and has an inclination angle of $\beta$ to the local
vertical above the magnetic point charge $M$. The line-of-sight component
$B_z$ of the magnetic field has an angle $(\alpha-\beta)$ to the radial
magnetic field component $B_r$. See details in Appendix A in Aschwanden
et al.~2012).}
\end{figure}
\clearpage

\begin{figure}
\plotone{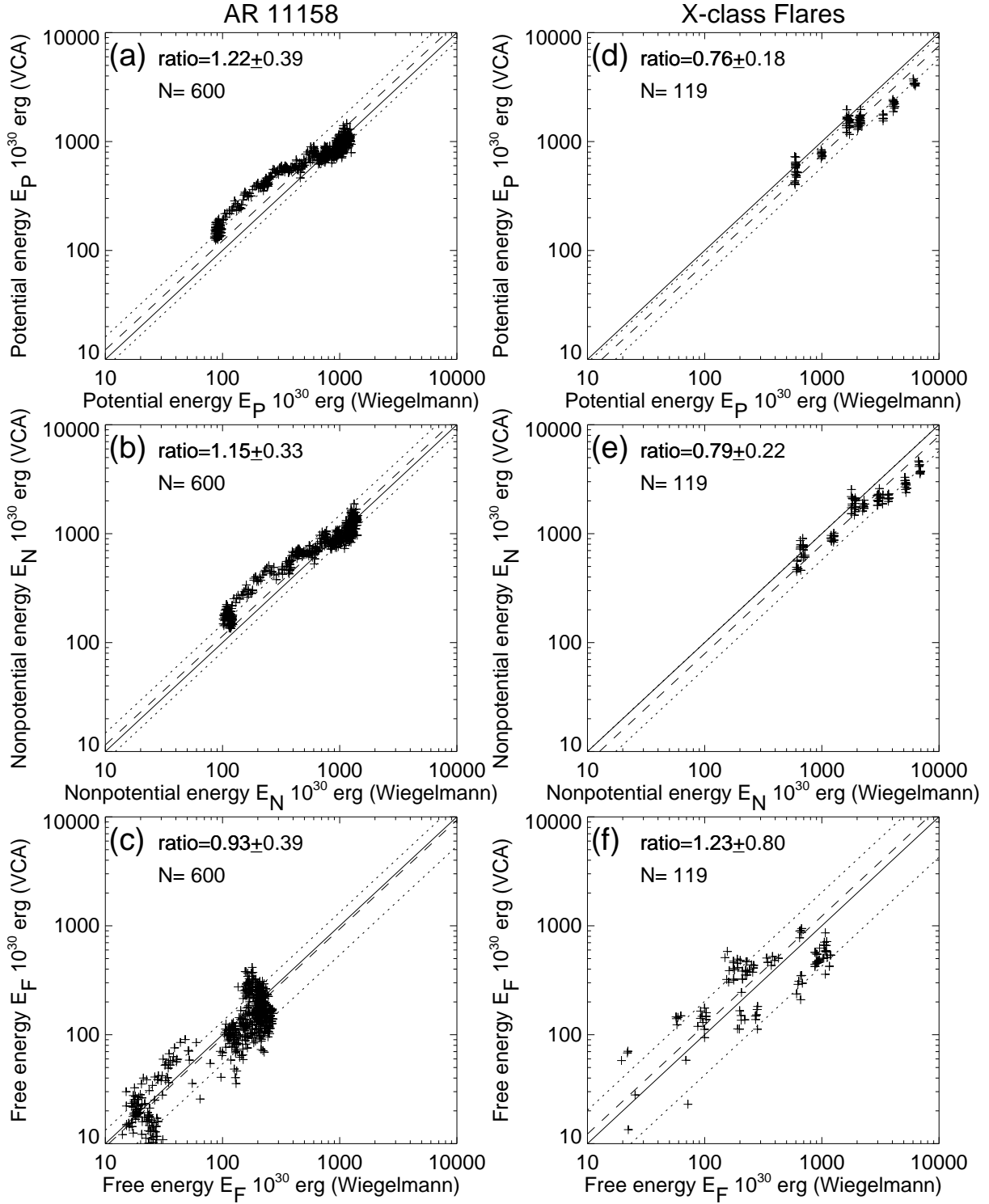}
\caption{Scatter plot of potential energies (top panels),
nonpotential energies (middle panels), and free energies (bottom panels)
measured with the VCA-NLFFF code (y-axis) and the Wiegelmann-NLFFF code
(x-axis) for 600 time steps of AR 11158 observed during 5 days
(left panels), and for 11 X-class flares (right panels). The diagonal
lines indicate equality (solid lines), mean ratios (dashed lines), and
standard deviations (dotted lines).}
\end{figure}
\clearpage

\begin{figure}
\plotone{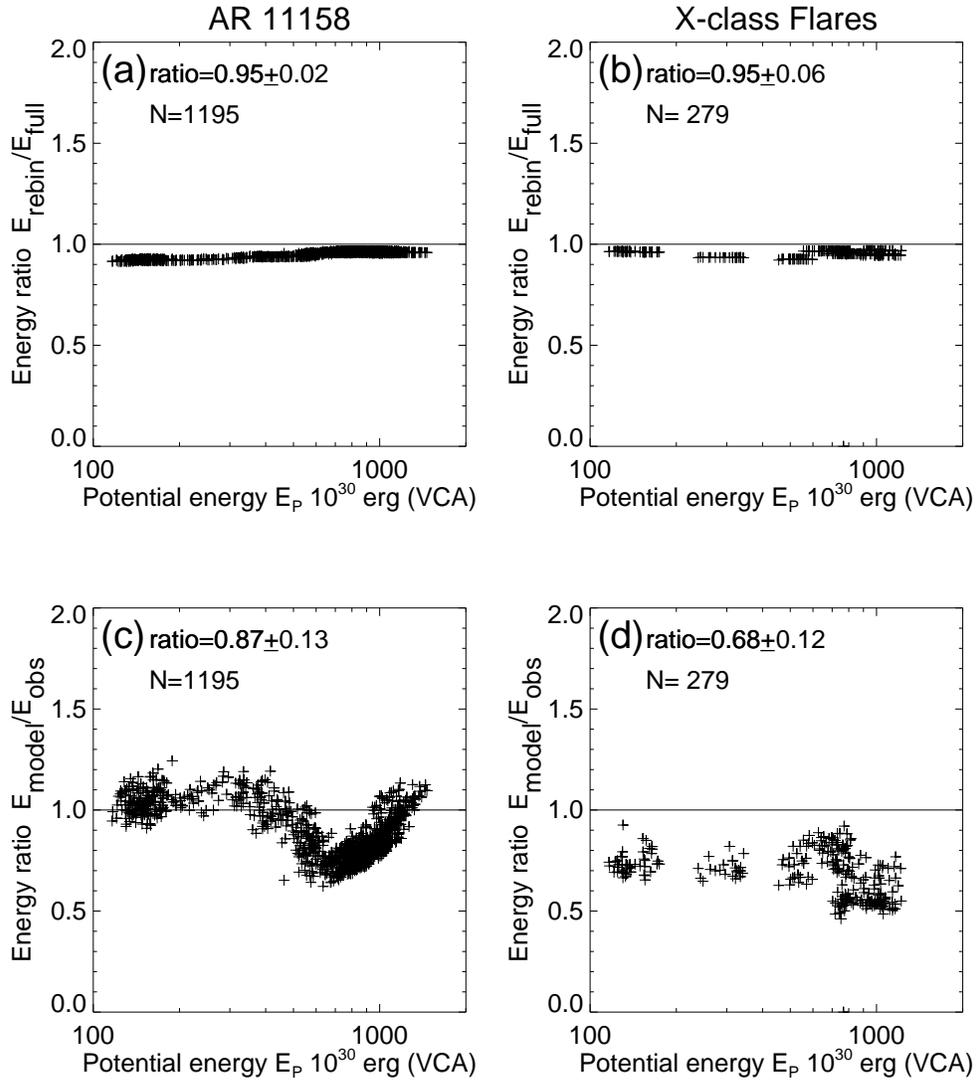}
\caption{Test of conservation of the potential energy due to rebinning
of the HMI magnetograms from $0.5\arcsec$ to $1.5\arcsec$ (top panels)
and due to the model representation of the magnetogram with a finite
number of (typically $n_{mag}=30$) magnetic source components
(bottom panels), for 1195 time steps of NOAA active region 11158
during 2011 Febr 12-17 (left panels), and 279 time steps of
11 X-class flares (right panels).}
\end{figure}
\clearpage

\begin{figure}
\plotone{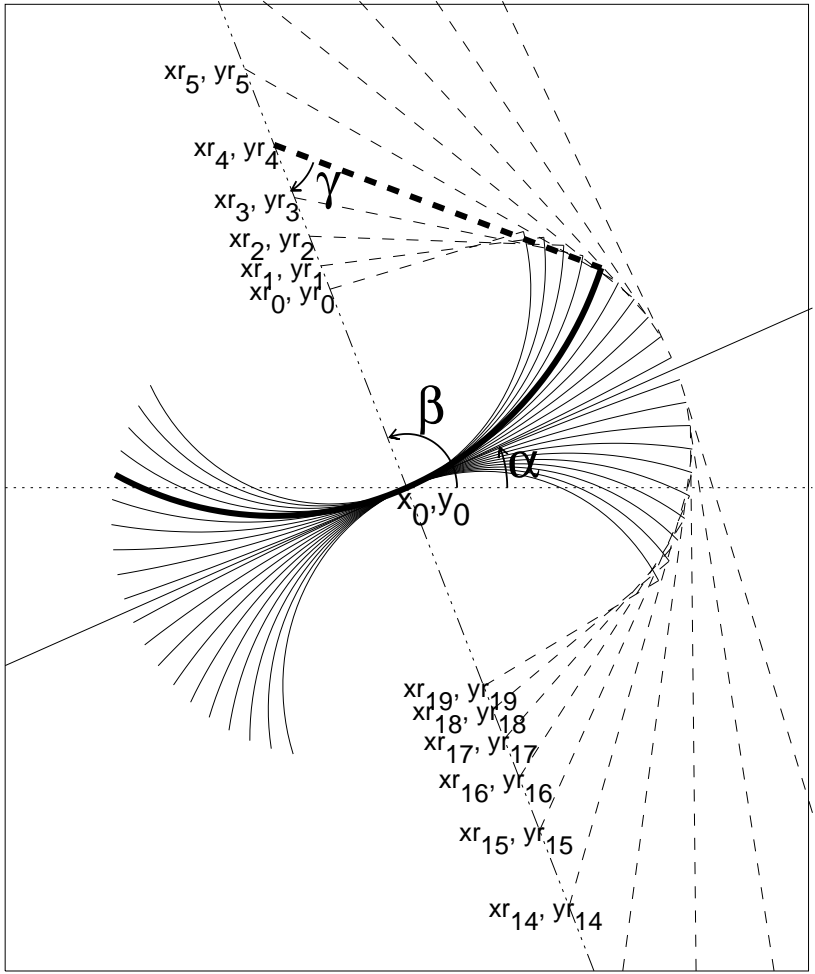}
\caption{The geometry of automated curvi-linear feature tracking 
is shown, starting at a local flux maximum location $(x_0,y_0$,
where the linear direction of the local ridge is measured
(angle $\alpha$) and a set of circular segments within a range
of curvature radii is fitted to the local ridge (thick linestyle). 
The locations $(xr_i, yr_i), i=0,...,19$ mark the centers of the 
curvature radii (Aschwanden, De Pontieu, and Katrukha 2013).}
\end{figure}

\begin{figure}
\plotone{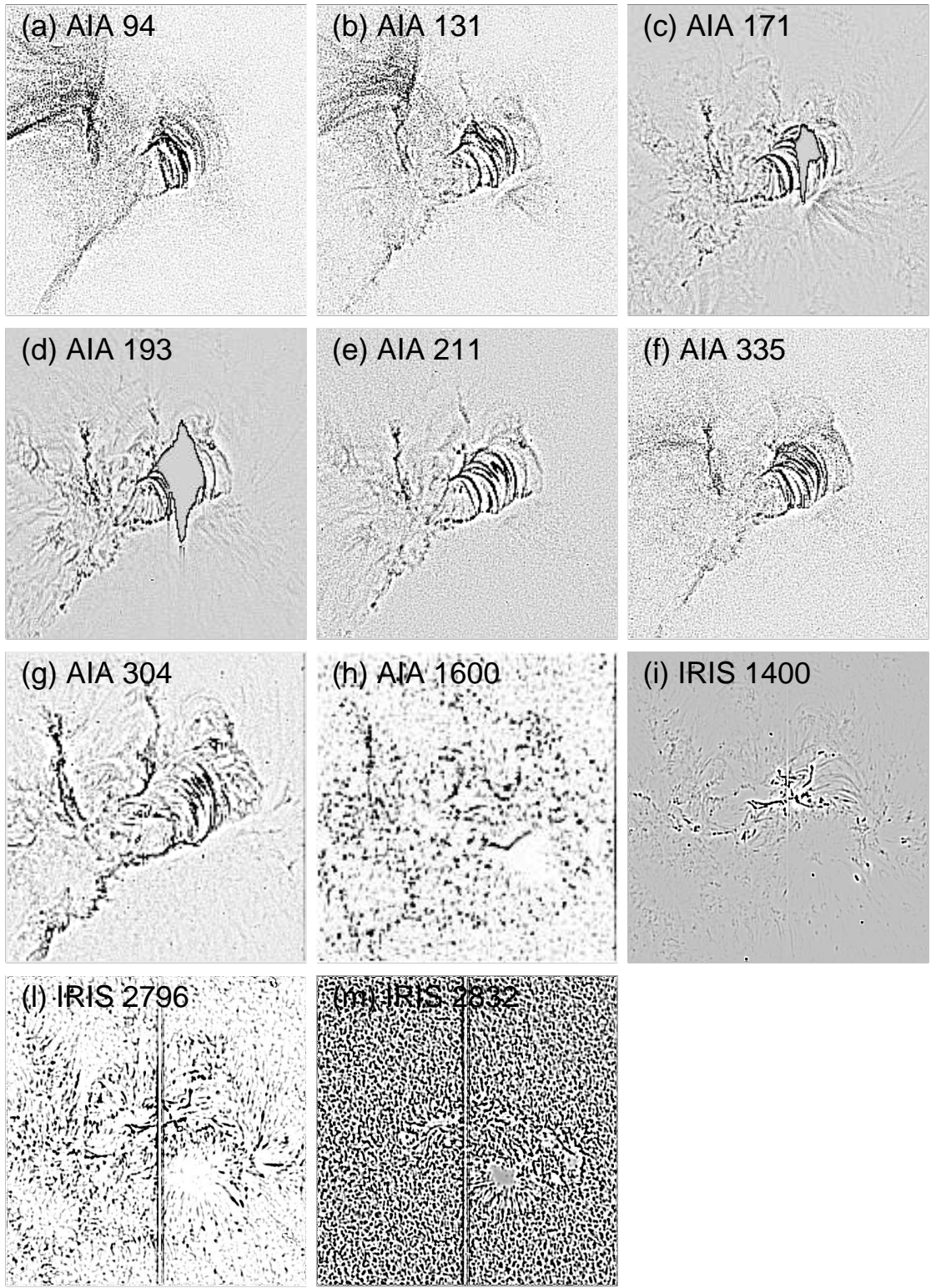}
\caption{Example of highpass-filtered images, observed on 2014 March 29, 
18:17 UT with AIA, and at 17:35 UT with IRIS. The images were obtained
with a lowpass filter of $n_{sm1}=1$ and a highpass filter of $n_{sm2}=3$.
Otherwise, respresentation is similar to Fig.~3.}
\end{figure}
\clearpage

\begin{figure}
\plotone{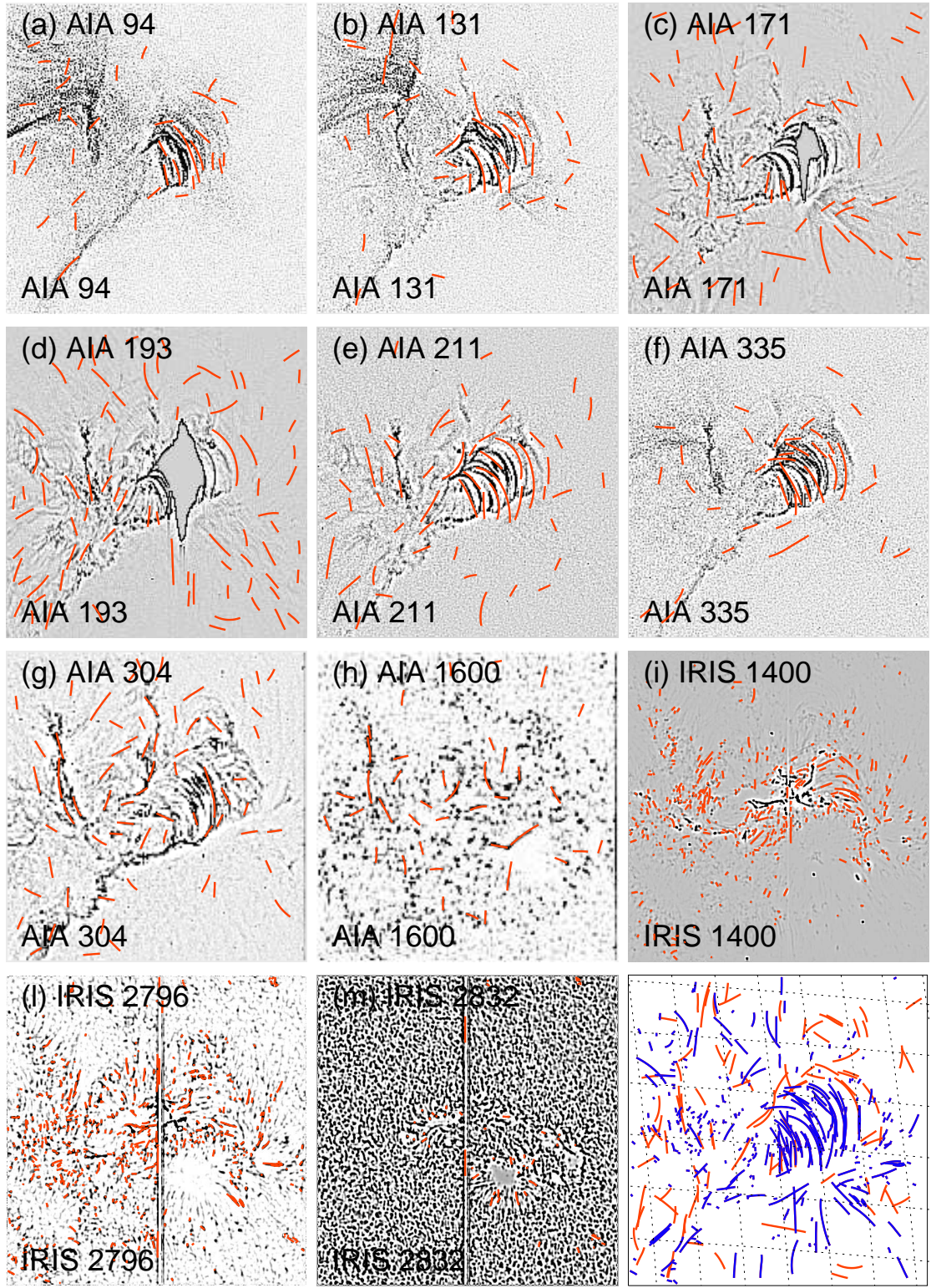}
\caption{Example of automated tracing of curvi-linear features, applied to
the highpass-filtered images shown in Fig.~9, observed on 2014 March 29, 
18:17 UT with AIA, and at 17:35 UT with IRIS. The combination of the
tracings in all wavelengths is shown in the bottom right panel. All
detected structures are marked with red color, while the fitted structures
(with a misalignment of $\le 20^\circ$) are marked with blue color.
Otherwise, respresentation is similar to Figs.~3 and 9.}
\end{figure}
\clearpage

\begin{figure}
\plotone{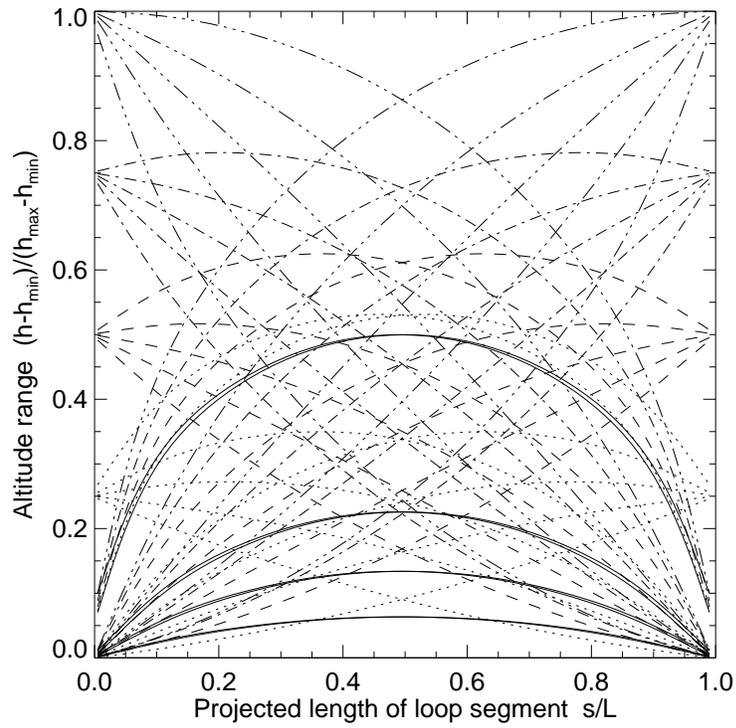}
\caption{Modeling of the altitude coordinate $h(s)$ of loop segments
observed in 2D, using a selected subset of circular geometries in the
loop plane. The loop segments are all circular, but have different
altitudes for the footpoints and apices, and different angular ranges
of a semi circle. The loop segments are shown here with a normalized
projected length $s/L$ (x-axis) and a normalized height range
$(h-h_{min})/(h_{max}-h_{min})$ (y-axis).}
\end{figure}
\clearpage

\begin{figure}
\plotone{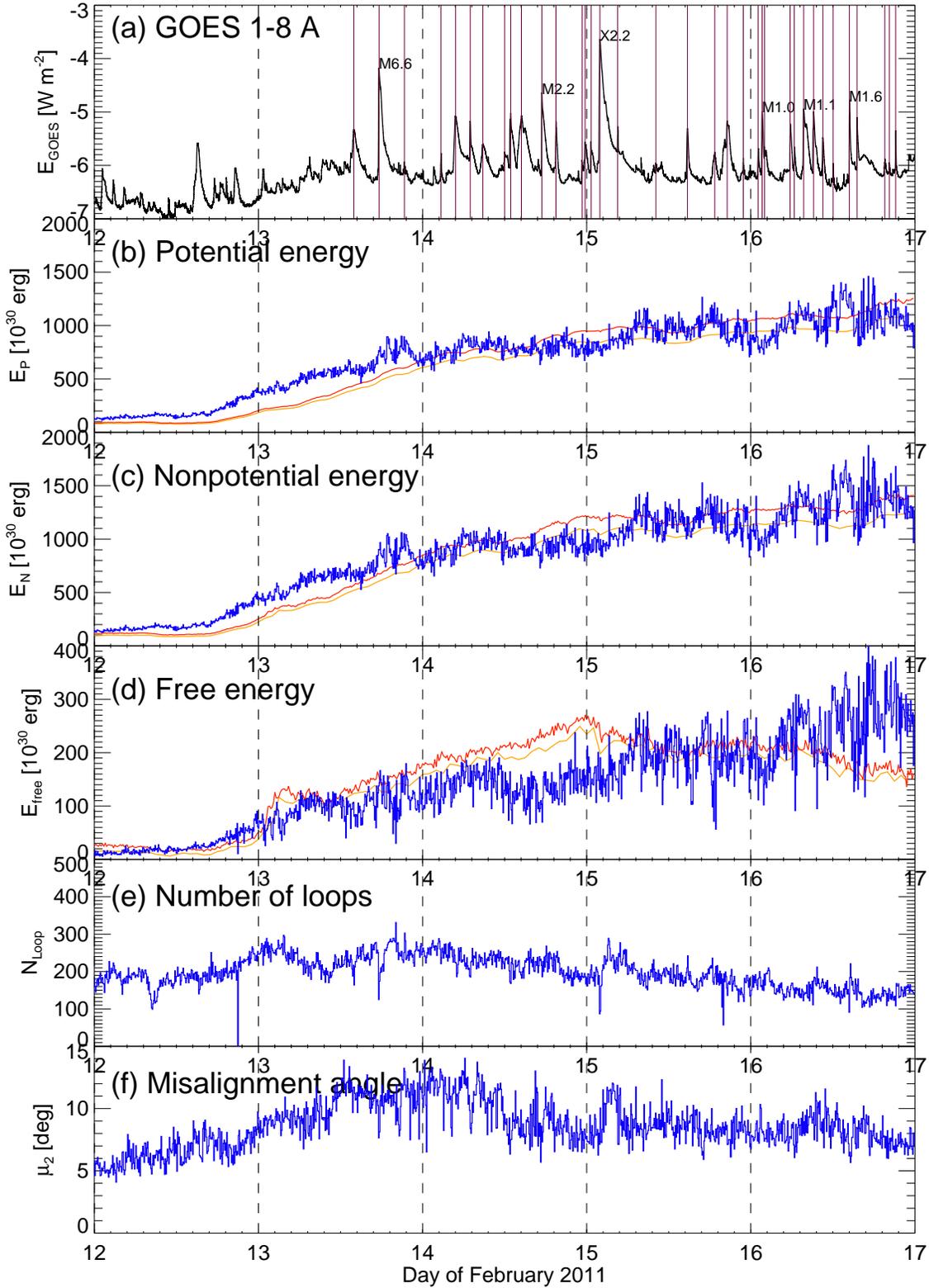}
\caption{Time evolution of magnetic energies of AR 11158
during 2011 Feb 12 to 17: (a) GOES 1-8 \ang\ flux, with GOES C-, M-,
and X-class flares indicated with purple vertical lines;
(b) Potential field energy $E_P$;
(c) Nonpotential energy $E_N$;
(d) Free energy $E_{free}=E_{N}-E_P$;
(e) The number of fitted loops $N_{loop}$;
(f): the 2D misalignment angle $\mu_2$ of the best fit. The color code
indicates forward-fitting of traced loops with the VCA-NLFFF code in
6-min time intervals (blue), the W-NLFFF code in 12-min intervals
(red) and 1-hr intervals (orange; Sun et al.~2012a).}
\end{figure}
\clearpage

\begin{figure}
\plotone{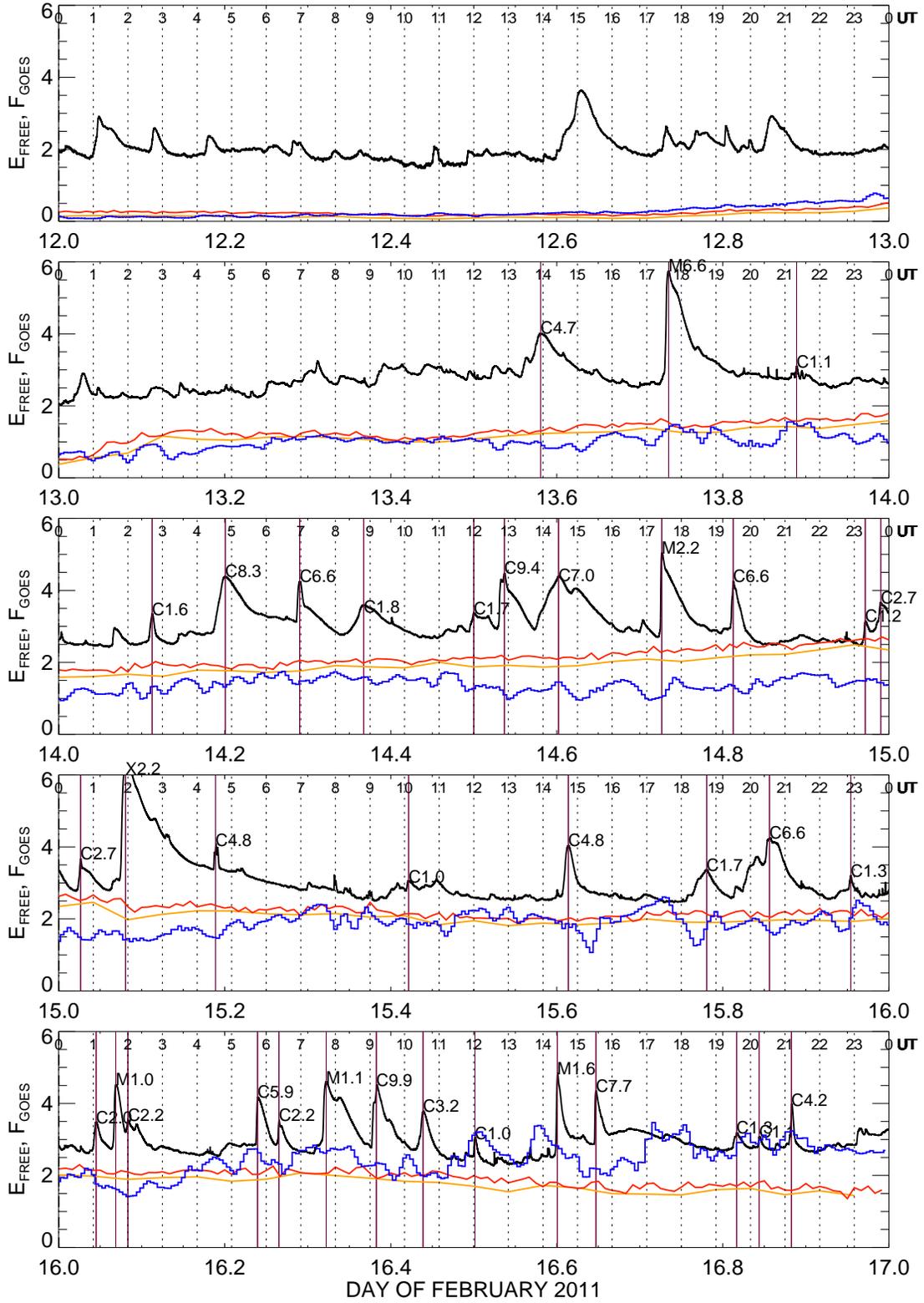}
\caption{Expanded (3-point median) time profiles of the 
(logarithmic) GOES 1-8 \ang\ flux (black,
arbitrary units), the free energies computed with the W-NLFFF code
in 12-min intervals (red) and 1-hr intervals (orange; Sun et al.~2012a),
with the VCA-NLFFF code in 6-min intervals
(blue). The times of 36 GOES C-,M-, and X-class flares occurring in
AR 11158 are indicated with vertical purple lines and labeled with
the GOES class. Each panel represents a consecutive day from
2011 Feb 12 to 17.}
\end{figure}
\clearpage

\begin{figure}
\plotone{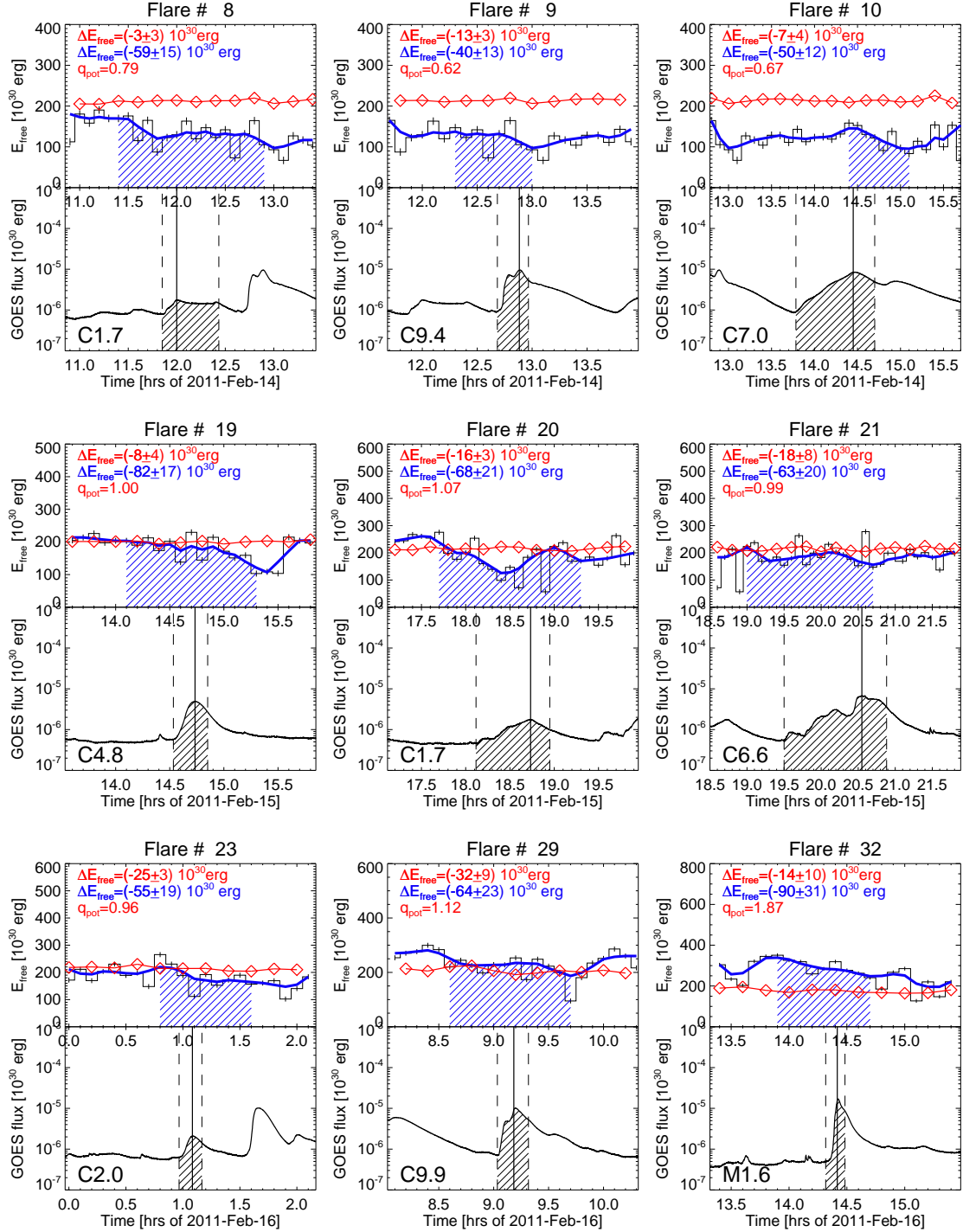}
\caption{The time evolution of the free energy $E_{free}(t)$ is shown
for 9 flare events occurring in AR 11158 during 2011 Feb 12-17. The 9 events
were selected by the largest significance of energy decreases $\Delta E_{free}$
during the GOES flare duration, including a time margin $\pm 0.5$ hours before
and after the flare. The free energy is determined with the W-NLFFF code
(red diamonds), and independently with the VCA-NLFFF code (histogram with
error bars), rendered as 3-point median (blue curve). The time interval
with energy decrease is marked with a blue hatched area. The GOES flux is
shown on a logarithmic scale. The time interval between start and end
of the flare is marked with a hatched (black) area.}
\end{figure}
\clearpage

\begin{figure}
\plotone{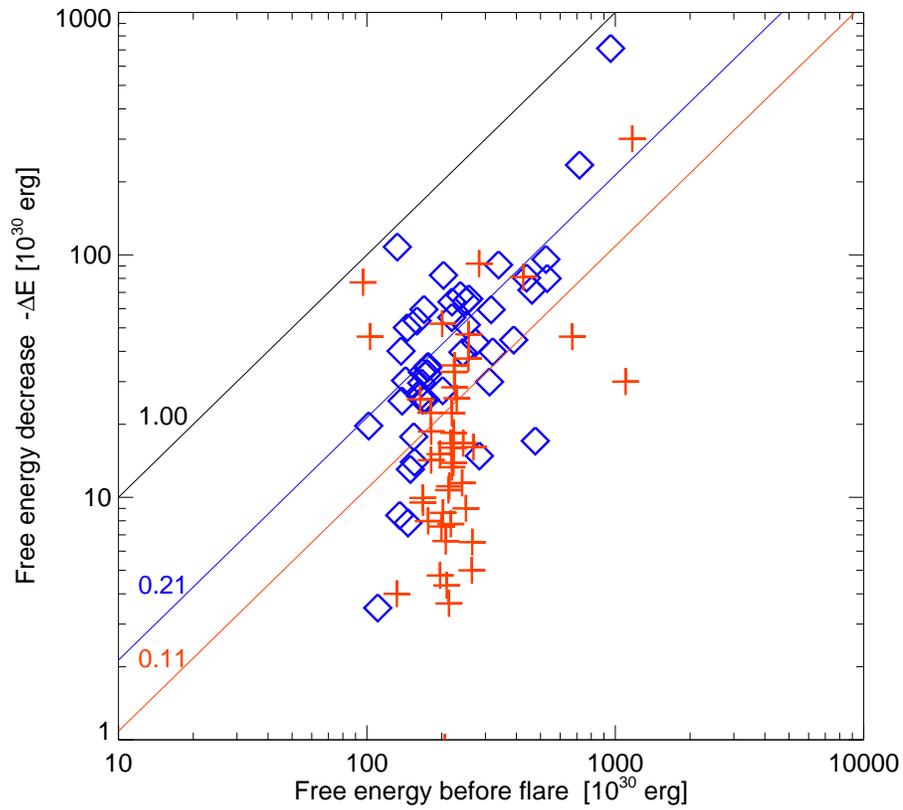}
\caption{Decreases of the free energy decrease during 36 flares of 
AR 11158 and 11 X-class flares
are plotted versus the free energy before the flare, as measured with
the VCA-NLFFF code (blue diamonds) and the W-NLFFF code (red crosses).
Note that VCA-NLFFF yields an energy decrease from the preflare free
energy by 21\% (in the (logarithmic average), while the W-NLFFF code 
detects about half of this value (11\%).}
\end{figure}
\clearpage

\begin{figure}
\plotone{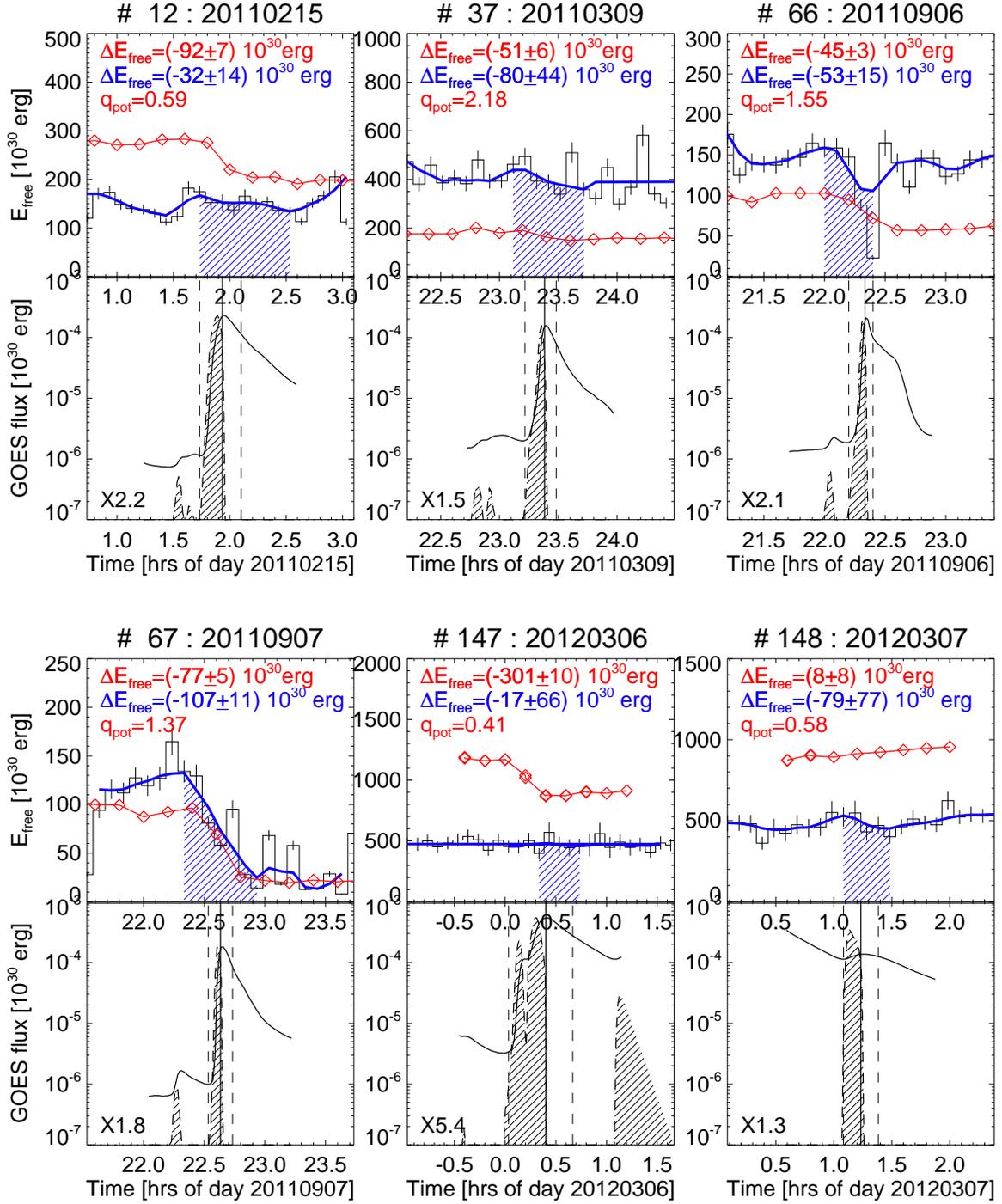}
\caption{For 6 X-class flare events we show the evolution of the free
energy $E_{free}(t)$ as determined with the VCA-NLFFF code (blue curve
and black histogram) and with the W-NLFFF code (red curve), along with
the GOES 1-8 \ang\ light curve. The time interval with decreases in the
free energy are marked with a blue hatched area, and the GOES time
derivative is marked with a black hatched area.}
\end{figure}
\clearpage

\begin{figure}
\plotone{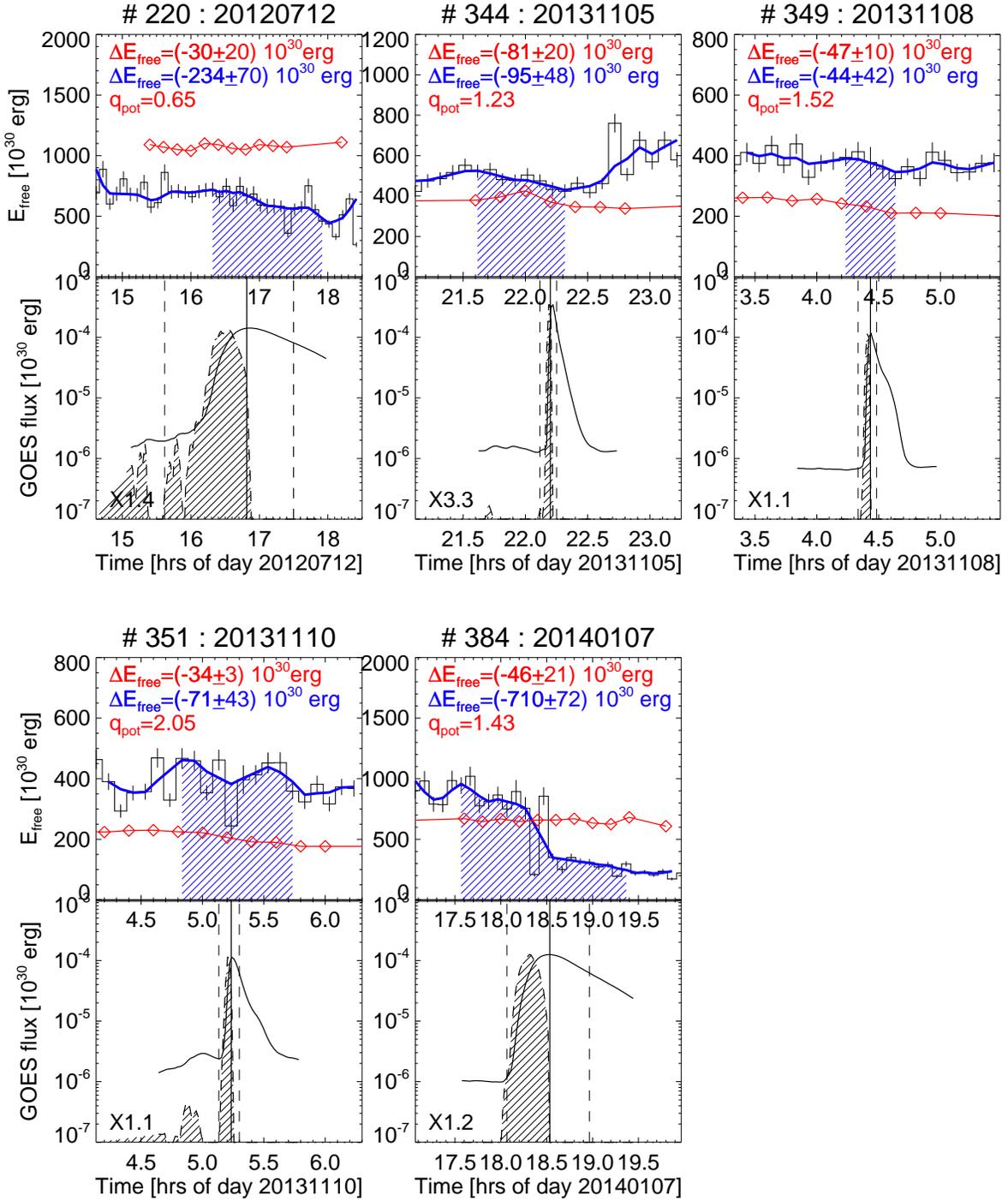}
\caption{Continuation of Fig.~16 with 5 additional X-class flares.}
\end{figure}
\clearpage

\begin{figure}
\plotone{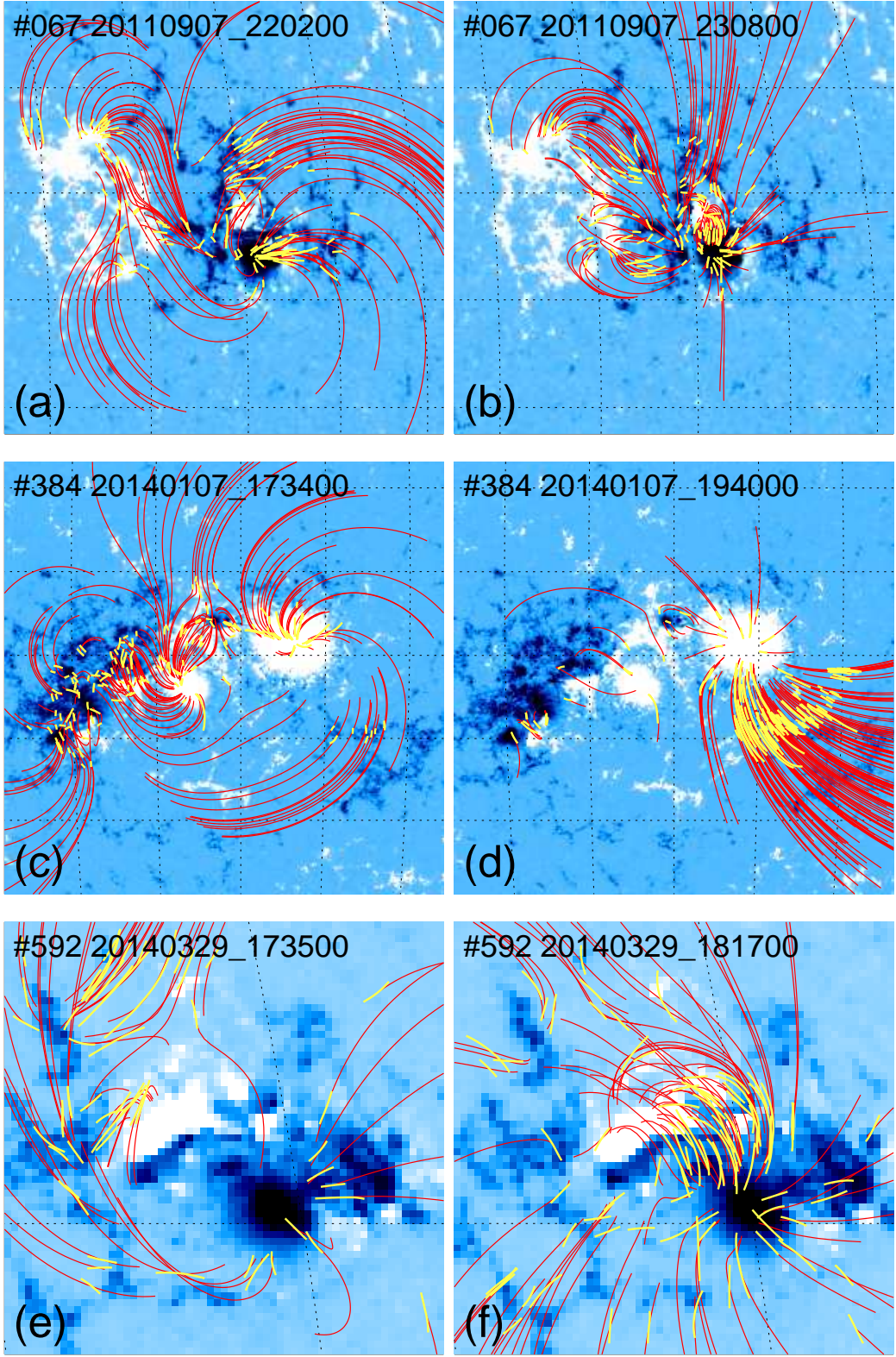}
\caption{The evolution of the three flares with the most pronounced
decrease in free energy are presented at flare start (left panels) 
and at flare end (right panels). The blue background image represents
the $B_z$ HMI magnetogram, the yellow curves present the automatically
traced loops, and the red curves represent the best-fit magnetic field
based on the VCA-NLFFF code, where only field lines that intersect with
the midpoint of the traced loops are shown.}
\end{figure}
\clearpage

\begin{figure}
\plotone{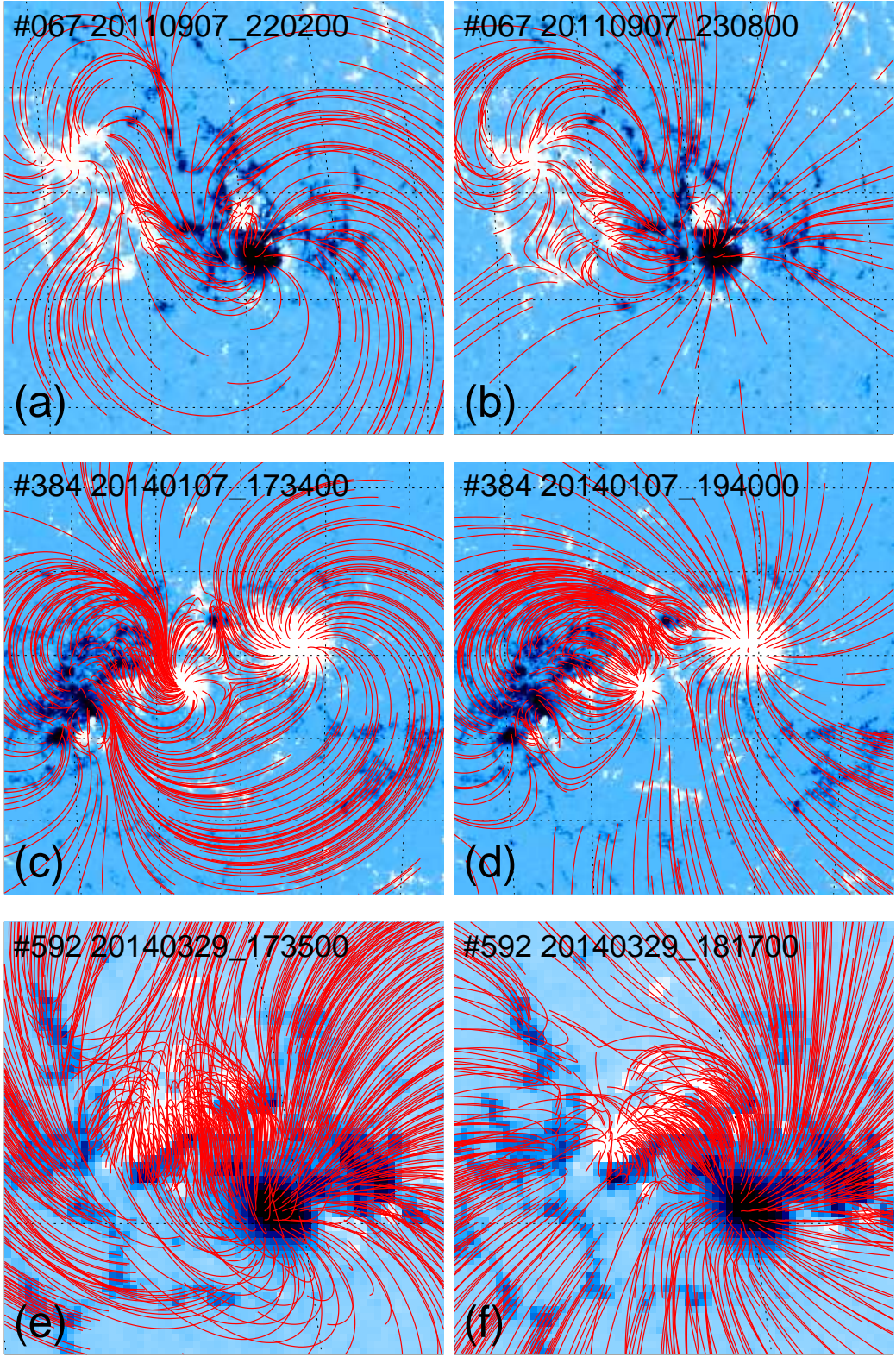}
\caption{The evolution of the nonlinear force-free magnetic field
before the flare (left panels) and after the flare (right panels). 
The blue background image represents the $B_z$ HMI magnetogram, 
while the red curves represent the best-fit magnetic field lines
computed with the VCA-NLFFF code, selected with footpoints in a
$50 \times 50$ grid with field strengths of $|B_z| > 100$ G.}
\end{figure}
\clearpage

\begin{figure}
\plotone{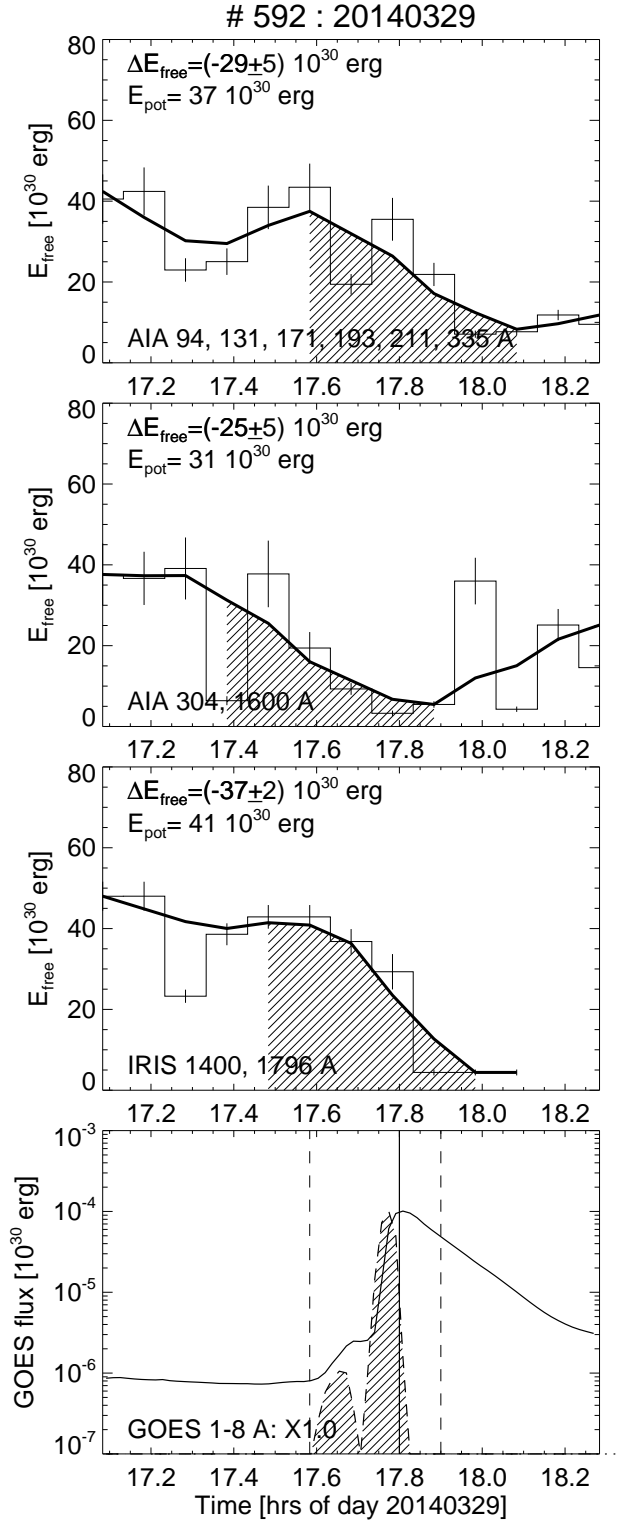}
\caption{Time evolution of the free energy as measured from
AIA in EUV wavelengths (top panel), from AIA UV wavelengths
(second panel), from IRIS UV wavelengths (third panel),
along with the GOES 1-8 \ang\ flux (solid linestyle in bottom panel)
and GOES time derivative (dashed linestyle in bottom panel).
The start, peak, and end time of the GOES event is indicated
with vertical lines.}
\end{figure}
\clearpage

\begin{figure}
\plotone{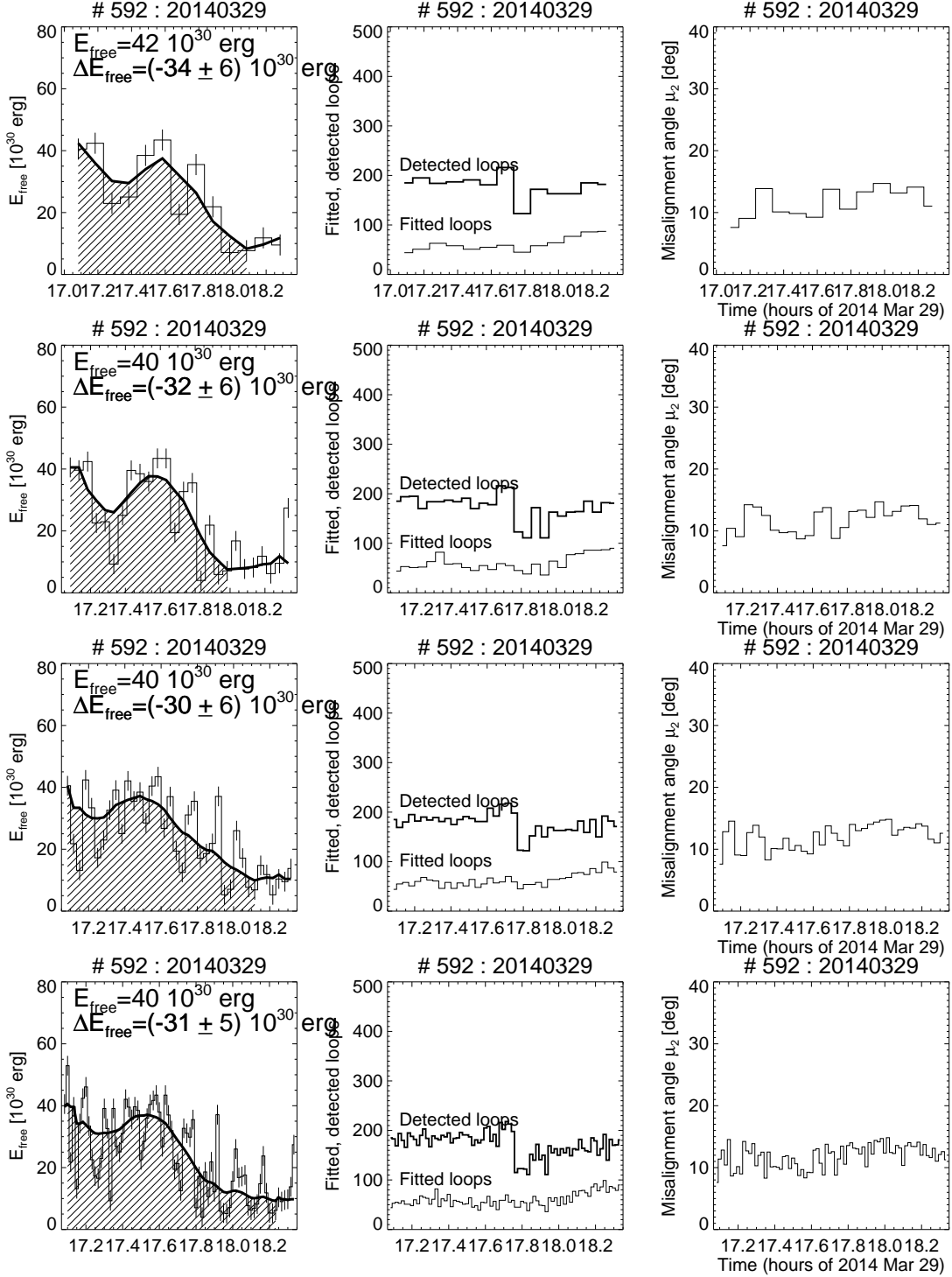}
\caption{Parametric study of the temporal evolution of the free energy
$E_{free}(t)$ (left column), the number of detected and fitted loop
segments (middle column), and the best-fit misalignment angle
$\mu_2(t)$ (right column), as a function of different time resolutions:
$t_{cadence}=6$ min (top), 
$t_{cadence}=2$ min (second row), 
$t_{cadence}=3$ min (third row), and 
$t_{cadence}=1$ min (bottom row).  
Note that the smoothed (3-point median) time profile is invariant to 
the time resolution of the data.}
\end{figure}
\clearpage

\begin{figure}
\plotone{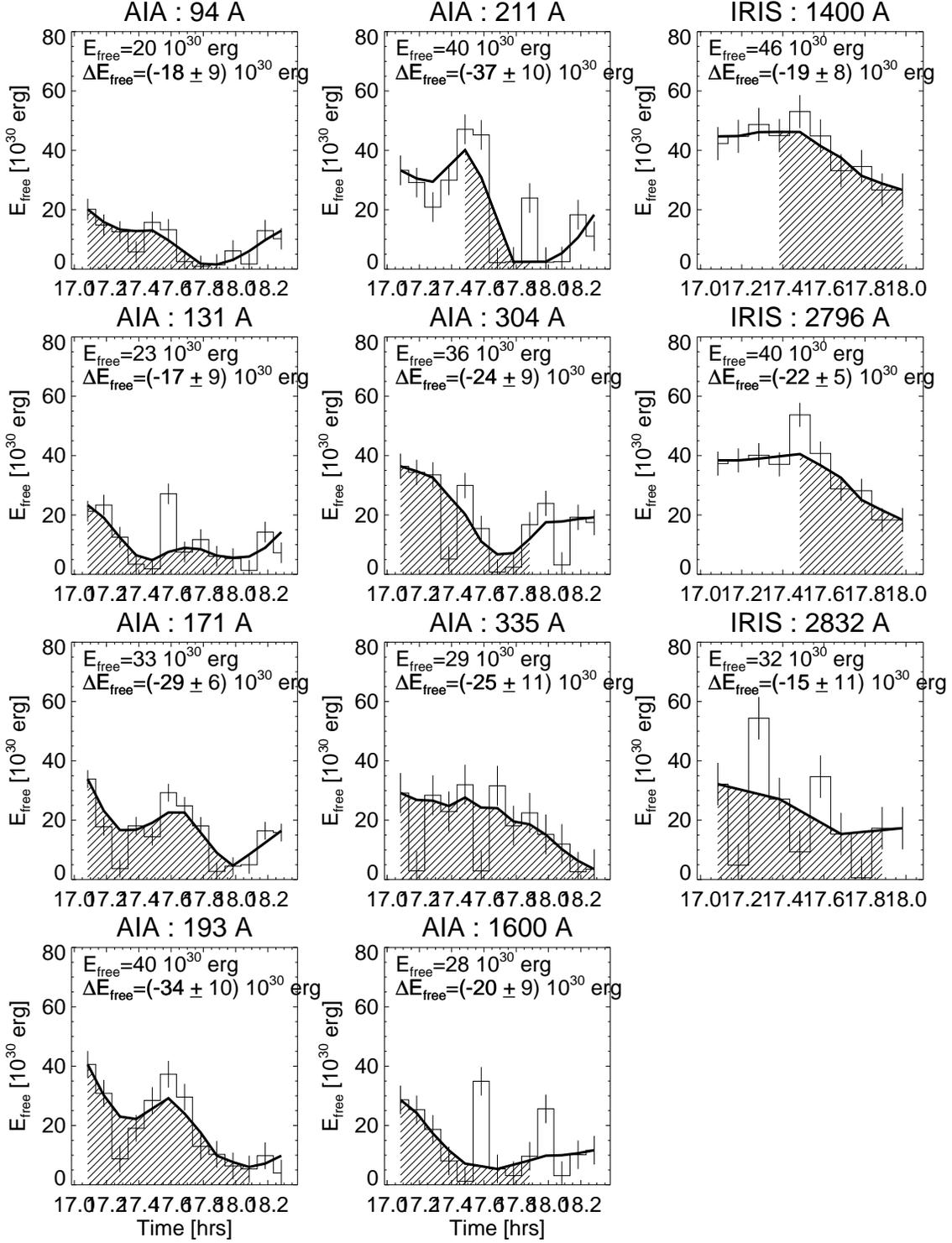}
\caption{Parametric study of the temporal evolution of the free energy
$E_{free}(t)$ as a function of time for each wavelength separately,
from AIA (left and middle column) and IRIS (right column).}
\end{figure}
\clearpage

\begin{figure}
\plotone{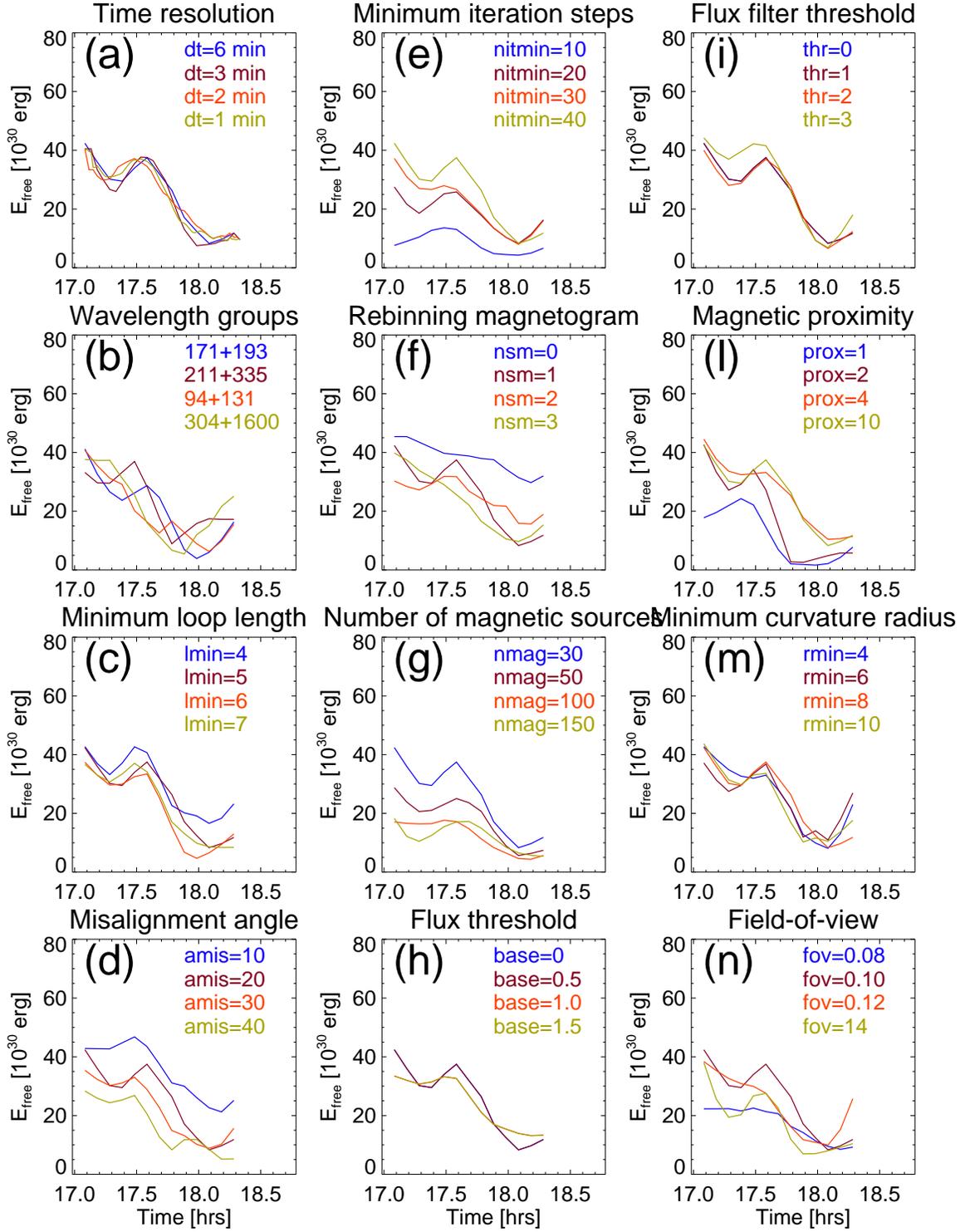}
\caption{Parametric study of the temporal evolution of the free energy
$E_{free}(t)$ as a function of energy, varying each of the 12 control 
parameters of the VCA-NLFFF code. Each panel shows the time profile 
$E_{free}(t)$ for 4 different values of each control parameter.}
\end{figure}
\clearpage

\end{document}